\input lanlmac
\def\href#1#2{{#2}}

\input epsf.tex

\overfullrule=0mm

\newcount\figno
\figno=0
\def\fig#1#2#3{
\par\begingroup\parindent=0pt\leftskip=1cm\rightskip=1cm\parindent=0pt
\baselineskip=11pt
\global\advance\figno by 1
\midinsert
\epsfxsize=#3
\centerline{\epsfbox{#2}}
\vskip 12pt
{\bf Fig.\ \the\figno:} #1\par
\endinsert\endgroup\par
}
\def\figlabel#1{\xdef#1{\the\figno}}
\def\encadremath#1{\vbox{\hrule\hbox{\vrule\kern8pt\vbox{\kern8pt
\hbox{$\displaystyle #1$}\kern8pt}
\kern8pt\vrule}\hrule}}


\magnification=\magstep1
\baselineskip=12pt
\hsize=6.3truein
\vsize=8.7truein
 at 8truept
 at 8truept
 at 10truept

\vbox{\hfill IPhT-t11/148}
\bigskip
\bigskip
\font\bigrm=cmr12 at 14pt \centerline{\bigrm A recursive approach to the
O(n) model on random maps}
\bigskip
\centerline{\bigrm via nested loops}

\bigskip\bigskip

\centerline{G. Borot, J. Bouttier and E. Guitter}
  \smallskip
  \centerline{Institut de Physique Th\'eorique}
  \centerline{CEA, IPhT, F-91191 Gif-sur-Yvette, France}
  \centerline{CNRS, URA 2306}
\centerline{\tt gaetan.borot@unige.ch}
\centerline{\tt jeremie.bouttier@cea.fr}
\centerline{\tt emmanuel.guitter@cea.fr}

  \bigskip

     \bigskip\bigskip

     \centerline{\bf Abstract}
     \smallskip
     {\narrower\noindent

We consider the $O(n)$ loop model on tetravalent maps and show how to rephrase
it into a model of bipartite maps without loops. This follows from a 
combinatorial decomposition that consists in cutting the $O(n)$ model 
configurations along their loops so that each elementary piece is a map 
that may have arbitrary even face degrees. In the induced statistics, these 
maps are drawn according to a Boltzmann distribution whose parameters (the 
face weights) are determined by a fixed point condition. In particular, we 
show that the dense and dilute critical points of the $O(n)$ model correspond 
to bipartite maps with large faces (i.e. whose degree distribution has a fat 
tail). The re-expression of the fixed point condition in terms of linear 
integral equations allows us to explore the phase diagram of the model. 
In particular, we determine this phase diagram exactly for the simplest 
version of the model where the loops are ``rigid''. Several generalizations 
of the model are discussed.
\par}

     \bigskip

\nref\TutteCPM{W.T. Tutte, {\it A Census of Planar Maps}, Canad. J. of Math.
{\bf 15} (1963) 249-271.}
\nref\GouldJack{See for instance Section 2.9 of: I.P. Goulden and D.M.
Jackson, {\it Combinatorial Enumeration}, John Wiley \& Sons, New
York (1983), republished by Dover, New York (2004), and references
therein.}
\nref\BIPZ{E. Br\'ezin, C. Itzykson, G. Parisi and J.-B. Zuber, {\it Planar
Diagrams}, Comm. Math. Phys. {\bf 59} (1978) 35-51.}
\nref\Schae{G. Schaeffer, {\it Conjugaison d'arbres
et cartes combinatoires al\'eatoires}, PhD Thesis, Universit\'e
Bordeaux I (1998).}
\nref\MierRev{See for instance: G. Miermont, {\it Random maps and their
scaling limits}, in C. Bandt, P. M\"orters, M. Z\"ahle (Eds.),
Proceedings of the conference Fractal Geometry and Stochastics IV,
Greifswald (2008),
Progress in Probability, Vol.\ {\bf 61}, 197-224, Birkha\"user (2009), and
references therein.}
\nref\DGZ{See for instance: P. Di Francesco, P. Ginsparg and
J. Zinn--Justin, {\it 2D Gravity and Random Matrices},
Physics Reports {\bf 254} (1995) 1-131,	arXiv:hep-th/9306153, 
and references therein.}
\nref\DK{B. Duplantier and I. Kostov, {\it Conformal spectra of polymers on
a random surface}, Phys. Rev. Lett. {\bf 61} (1988) 1433-1437.}
\nref\Kost{I. Kostov, {\it $O(n)$ vector model on a planar random lattice:
spectrum of anomalous dimensions}, Mod. Phys. Lett. {\bf 4} (1989) 217-226.}
\nref\KostSta{I. Kostov and M. Staudacher, {\it Multicritical Phases of the 
O(n) Model on a Random Lattice}, Nucl. Phys. {\bf B384} (1992) 459-483,
arXiv:hep-th/9203030.}
\nref\EZJ{B. Eynard and J. Zinn--Justin, {\it The $O(n)$ model on a random
surface: critical points and large order behaviour}, Nucl. Phys. {\bf B386}
(1992) 558-591, arXiv:hep-th/9204082.}
\nref\EK{B. Eynard and C. Kristjansen, {\it Exact solution of the $O(n)$
model on a random lattice}, Nucl. Phys. {\bf B455} (1995) 577-618, 
arXiv:hep-th/9506193.}
\nref\EKmore{B. Eynard and C. Kristjansen, {\it More on the exact solution
of the $O(n)$ model on a random lattice and an investigation of the
case $|n|>2$}, Nucl. Phys. {\bf B466} (1996) 463-487,
arXiv:hep-th/9512052.}
\nref\LGM{J.-F. Le Gall and G. Miermont, {\it Scaling limits of random planar 
maps with large faces}, Ann. Probab. {\bf 39(1)} (2011) 1-69, arXiv:0907.3262 
[math.PR].}
\nref\BE{G. Borot and B. Eynard, {\it Enumeration of maps with self avoiding 
loops and the O(n) model on random lattices of all topologies}, 
J. Stat. Mech. (2011) P01010, arXiv:0910.5896.}
\nref\MOB{J. Bouttier, P. Di Francesco and E. Guitter. {\it
Planar maps as labeled mobiles},
Elec. Jour. of Combinatorics {\bf 11} (2004) R69, arXiv:math.CO/0405099.}
\nref\HANKEL{J. Bouttier and E. Guitter, {\it Planar maps and continued 
fractions}, arXiv:1007.0419.}
\nref\Dieudonne{J. Dieudonn\'e, {\it Calcul infinit\'esimal}, 2nd Edition,
Hermann, Paris (1980) Chapter IV, Section 2.}
\nref\FS{P. Flajolet and R. Sedgewick, {\it Analytic Combinatorics}, Cambridge 
University Press (2009).}
\nref\HLE{A.V. Manzhirov and A.D. Polyanin, {\it Handbook of Integral 
Equations}, Chapman \& Hall/CRC (2008), p 625.}
\nref\Tric{F.G. Tricomi, {\it Integral Equations}, Interscience Publishers
(1957).}
\nref\GBThese{G.~Borot, PhD Thesis, Universit\'{e} d'Orsay (2011),
arXiv:1110.1493.}
\nref\BOUKA{D. Boulatov and V. Kazakov, {\it The Ising model
on a random planar lattice: the structure of the phase 
transition and the exact critical exponents}, Phys. Lett. {\bf B186} (1987)
379-384.}
\nref\Nien{B. Nienhuis, {\it Phase transitions and critical phenomena},
Vol.\ 11, eds.\ C.\ Domb and J.L.\ Lebowitz, Academic Press (1987).}
\nref\EKbis{B. Eynard and C. Kristjansen, {\it An Iterative Solution 
of the Three-colour Problem on a Random Lattice}, Nucl. Phys. {\bf B516} 
(1998) 529-542, arXiv:cond-mat/9710199.}
\nref\WangGuo{Z.X.~Wang and D.R.~Guo, {\it Special functions}, World Scientific, reprinted 2010.}

\newsec{Introduction and main results}

\subsec{General introduction}
Planar maps, which are proper embeddings of graphs in the two-dimensional
sphere, are fundamental mathematical objects whose combinatorics raises
many beautiful enumeration problems, first addressed by Tutte in the
60's \TutteCPM. Maps are also widely used in physics as discrete models for
fluctuating surfaces or interfaces in various contexts, ranging from soft
matter physics (e.g. biological membranes) to high energy physics (e.g. string
theory). Most problems involve ``random maps'', i.e. consist in the study of
some particular statistical ensemble of maps, distributed according to some
prescribed law. Interesting scaling limits may be reached when
one considers ensembles of large random maps, giving rise to nice universal
probabilistic objects.
Several map enumeration techniques were developed over the years,
which we may classify in three categories: Tutte's original recursive
decomposition method \GouldJack, the technique of matrix integrals \BIPZ\ and,
more recently, the bijective method \Schae\ where maps are coded by tree-like
objects.

So far, the most advanced results were obtained for ensembles of maps with
a simple control on the degree of, say the faces of the maps. Examples are
ensembles of triangulations (maps with faces of degree $3$ only) or of
quadrangulations (degree $4$) where the total number of faces in the map is
either fixed or governed by some Boltzmann distribution. The statistics of
these maps is now well understood and many
exact enumeration results were obtained within the framework of each of the
three above enumeration techniques. Of particular interest is the scaling
limit of large maps with prescribed {\it bounded degrees}, which realizes
the so-called universality class of ``pure gravity'', and gives rise
to the ``Brownian map'', a probabilistic object with remarkable metric
properties \MierRev. As first recognized in physics, other universality classes
of maps may be reached upon equipping the maps with statistical models,
such as models of spins or particles, which present a large variety
of critical phenomena that modify the statistical properties of the
underlying maps.
This gives rise to a large variety of universality classes for maps, whose
understanding is the domain of the so-called ``two-dimensional quantum
gravity'' \DGZ. It is worth mentioning that most results in this domain rely
on matrix integral formulations of the models, so that the matrix integral
technique appears so far as the most powerful approach to explore new
universality classes.

A particular important class of models of maps equipped with statistical
models are the so-called $O(n)$ loop models which consist in having
maps endowed with configurations of closed self- and mutually-avoiding
loops drawn on their edges, each loop receiving the weight $n$.
A particular class of $O(n)$ loop models where loops visit only
vertices of degree $3$ (the degree of the other vertices being bounded)
was analyzed in details by use of matrix integral techniques
in [\xref\DK-\xref\EKmore]. There it was found that these models present
several phases with non-trivial universal scaling limits. We will briefly
recall the results of this analysis in Section 1.2 below.

It was recognized recently in \LGM\ that another simple way to escape from the
universality class of pure gravity consists in considering
``maps with large faces'', i.e. ensemble of maps where the degree of
the faces is {\it unbounded} and properly controlled so that faces with
arbitrarily large degree persist in the scaling limit. Using now the bijective
method, it was shown that such ensembles may give rise to new
probabilistic objects corresponding to maps coded by stable trees.
It was then proposed that these probabilistic objects may also describe
the scaling limit of $O(n)$ loop models in some of their phases.

The purpose of this paper is to show that this is indeed the case.
More precisely, we show that a number of $O(n)$ loop models may be
bijectively transformed into models of maps without loops and with a simple
control on the degree of their faces. This control involves some
degree-dependent face weights whose value is fixed by some appropriate
consistency relation in the form of a fixed point condition depending on $n$. 
This implies that the
possible scaling limits of the $O(n)$ loop models necessarily match
the scaling limits of maps controlled simply by their face degrees.
These include the pure gravity universality class as well as the new
universality classes of Ref.~\LGM\ for maps with large faces, which are
observed in the dense and dilute phases of the $O(n)$ loop models.

\subsec{Previously solved $O(n)$ loop models}

As a reminder, we briefly describe a few of the known results about $O(n)$
models on random maps, as previously studied in [\xref\DK-\xref\EKmore] by
matrix integral techniques. We will not give very precise definitions here as
these models are not those that we will study in the remainder of the paper.
The $O(n)$ loop model studied in [\xref\DK-\xref\EKmore] consists in having
self- and mutually-avoiding loops drawn on random maps so as to visit only
vertices of degree $3$, the vertices not visited by loops having arbitrary
(but bounded) degrees at least $3$. 
A vertex visited by a loop receives the weight ${\tilde h}$ while
a vertex not visited by a loop receives the weight ${\tilde g}_k$ if it is
$k$-valent ($k\geq 3$). Each loop receives in addition the weight $n$.
We define ${\tilde F}_p^{\rm loop}$ as the corresponding generating 
function for loop configurations on rooted maps, where the 
root vertex is not visited and has degree $p$ (by convention, 
the root vertex is unweighted and we set ${\tilde F}_0^{\rm loop}=1$). 
This corresponds to having a dual map with a boundary of length $p$.
We may gather the family $({\tilde F}_p^{\rm loop})_{p\geq 0}$ into 
the resolvent
\eqn\resolv{{\tilde W}(\xi)=\sum_{p\geq 0} {{\tilde F}_p^{\rm loop}
\over \xi^{p+1}}\ ,}
well-defined for $\xi$ large enough. In the range of weights where the model
is well-defined, this formal series is in fact an analytic function on
${\bf C}\setminus[\gamma_-,\gamma_+]$, and has a discontinuity on some segment
$[\gamma_-,\gamma_+]$. Using this information in the recursive relation for
${\tilde F}_p^{\rm loop}$ obtained (in the spirit of the Tutte's recursive
method) by removing the root edge, it can be shown that ${\tilde W}(\xi)$ is
solution of a scalar non-local Riemann-Hilbert problem:
\eqn\RHprobW{
\forall \xi \in [\gamma_-,\gamma_+],\qquad {\tilde W}(\xi + {\rm i}0) +
{\tilde W}(\xi - {\rm i}0) + n{\tilde W}({\tilde h}^{-1} - \xi) =
\xi-\sum_{k \geq 3} {\tilde g}_k\,\xi^{k - 1}\ .}
This fixes uniquely ${\tilde W}(\xi)$ as well as $\gamma_-$ and $\gamma_+$,
from the requirement that ${\tilde W}(\xi) \sim 1/\xi$ when $\xi
\rightarrow \infty$, and that ${\tilde W}$ is holomorphic in
${\bf C}\setminus[\gamma_-,\gamma_+]$. A non-trivial critical point is
reached when $\gamma_+ \rightarrow (2{\tilde h})^{-1}$, which defines
a non-trivial critical surface ${\tilde h}={\tilde h}((\tilde g_k)_{k\geq 1})$.
The equation \RHprobW\ was first solved on this critical surface by Kostov
\Kost, who showed that ${\tilde W}$ may develop a singularity of the form:
\eqn\resolvasymp{{\tilde W}(\xi)\vert_{\rm sing.}\sim {\rm const.}
\left(\xi-{1\over 2{\tilde h}}\right)^{a-1} \quad {\rm for} \quad
\xi\to \left({1\over 2{\tilde h}}\right)^+\ ,}
with $a=2\pm b$, $\pi b= \arccos(n/2)$.
This singularity captures the large $p$ asymptotics of
${\tilde F}_p^{\rm loop}$ as
\eqn\asymresolv{{\tilde F}_p^{\rm loop}\sim {\rm const.} {(2{\tilde h})^{-p}
\over p^a}.}
The smallest value $a=2-b$ is observed inside the non-trivial critical surface
and describes the so-called {\it dense phase} of the $O(n)$ model, while
the largest value $a=2+b$ is observed only at some boundary of this
critical surface and describes the so-called {\it dilute phase}. Other
multicritical points, with $a = 2 \pm b + 2m$ $(m=1,2,\ldots)$, can also be
observed by tuning more and more coefficients ${\tilde g}_k$ \KostSta, but this
requires having some of these weights negative, which prevents interpreting
these multicritical points as proper probabilistic ensembles.
Eq.~\RHprobW\ was then solved in all generality (outside of the critical
surface) in \EK\ in terms of elliptic functions and of the Jacobi theta function
$\vartheta(\cdot,q)$, the non-trivial critical surface corresponding to the
limit where the nome $q$ goes to $1$ (or $0$ depending on the convention). 
In this limit, the theta function
degenerates into a trigonometric function, and one can recover the critical
exponents stated above. The study of the $O(n)$ model was extended recently
to maps of any topology in Ref.~\BE.

\subsec{Overview of the paper}
Let us start by giving a brief description of our results. Most of the
paper deals with $O(n)$ loop models defined on planar {\it tetravalent maps}
(maps with vertices of degree $4$ only). By definition, the loops are
self- and mutually-avoiding and each loop receives the weight $n$.
Different models are obtained by assigning different weights to the vertices
depending on whether they are visited by a loop or not. A special attention
will be paid to the simplest version of the model, the {\it rigid loop model},
where loop turns are forbidden at the tetravalent vertices.

Our analysis is simplified by first reformulating our loop model on
the dual maps, which are planar quadrangulations, and by then
extending them to quadrangulations with a boundary of arbitrary
(necessarily even) length.  Indeed, a simple characterization of the
universality class of the loop model at hand is via the large $p$
asymptotics of its generating (or partition) function $F_p^{\rm loop}$
in the presence of a boundary of length $2p$. It is expected to take the form
\eqn\asymgen{F_p^{\rm loop}\sim {\rm const.} {{\cal A}^p\over p^a}}
with a non-universal exponential growth factor ${\cal A}$ and a power law
decay factor $1/p^a$ involving a {\it universal exponent} $a$
in the range $3/2\leq a\leq 5/2$.
Assuming this asymptotic behavior, we show that, for a given $n$, at most four 
values of the exponent
$a$ may be observed, namely $a=3/2$, corresponding to a {\it subcritical}
model, $a=5/2$, corresponding to a {\it generic critical} model, and
\eqn\valagen{a=2\pm b, \quad \pi b= \arccos(n/2)}
corresponding to a {\it non-generic critical} model. This last behavior
requires that $n$ lies in the range $0<n<2$ (note that, for $n=0$, \valagen\
yields the pure gravity exponents $a=3/2$ and $a=5/2$). In the case of the 
rigid loop model, we solve the model exactly, thus establishing the
validity of \asymgen\ and providing a precise description of the phase diagram
which specifies the domain of physical parameters where each of the above
values of $a$ is observed.

The expressions \asymgen\ and \valagen\ are identical to those obtained 
for the $O(n)$ loop models discussed in Section 1.2. Here, they are 
obtained by different
means which rely on an equivalence of our models with models of bipartite
maps. More precisely, quadrangulations (with a boundary) endowed with loop
configurations may be coded bijectively by bipartite maps without loops but
with faces of arbitrary degree, each containing an $O(n)$ loop model
configuration of its own. This bijective decomposition allows, via a simple
substitution procedure,
to express $F_p^{\rm loop}$ in terms of the well-understood generating
function for bipartite maps with a boundary. As a consequence, we may
identify the
possible asymptotics of $F_p^{\rm loop}$ as those of ordinary maps with
possibly large faces, leading eventually to \asymgen\ and \valagen.

The paper is organized as follows. Section 2 is devoted to a combinatorial
study of the $O(n)$ loop model on tetravalent maps. After defining the
model in Section 2.1, we present in Section 2.2 a bijective decomposition 
which allows to code the configurations of our model in terms of their
{\it gasket}, which is a bipartite map with no loops but with {\it holes},
which are faces of arbitrary even degree, together with a content for each
hole. For consistency, the effective weights for the holes are shown
to obey some crucial {\it fixed point condition} which determines them
uniquely. This condition involves in particular some
ring generating function, which accounts for the configuration in the
immediate vicinity of a loop, and which we make explicit in Section 2.3.
Section 3 does not treat the $O(n)$ model itself, but discusses the
possible critical behaviors of bipartite maps, which we classify into 
subcritical, generic critical and non-generic critical. The first two
cases are discussed in Section 3.2 while a particular attention is paid to 
the non-generic critical behavior in Section 3.3. Section 3.4 discusses the
so-called resolvent of bipartite maps. 
Section 4 is devoted to the consequences of the fixed point condition
on the asymptotics of $F_p^{\rm loop}$, leading to \asymgen\ and \valagen\
above. We first concentrate on the rigid loop model in Section 4.1 before
addressing the general case in Section 4.2. Section 5 shows how to transform
the fixed point condition into a linear integral equation, both in
the rigid case (Section 5.1) and in the non-rigid case (Section 5.2). We
then analyze in Section 5.3 the solution of this equation as it captures
all the phase diagram of the $O(n)$ loop model. Section 6 presents a detailed
analysis of the rigid loop model. We first show in Section 6.1 how to transform
the fixed point condition into an equation for the resolvent of the model.
This equation is then solved, first along a line of non-generic critical
points in Section 6.2, then in all generality in Section 6.3. The
generic critical line is discussed in Section 6.4 and we summarize the 
resulting phase diagram in Section 6.5. Section 7 considers extensions
of our results to other classes of $O(n)$ loop models. We first show how
the relation \valagen\ may be modified by considering loops with
non-symmetric weights (Section 7.1) or with additional constraints
on their length (Section 7.2). We end our study by a discussion of
the wide class of $O(n)$ loop models defined on arbitrary even-valent maps
(with bounded degrees) and show that our results nicely extend to this
case. We gather our conclusions in Section 8.

\newsec{The $O(n)$ loop model on tetravalent maps: combinatorics}

\subsec{Definition of the model}

Our model of interest is the $O(n)$ loop model on planar tetravalent
maps defined as follows. By {\it loop} we mean an undirected simple
closed path on the map (visiting edges and vertices), also sometimes
called an undirected cycle.  A {\it loop configuration} is a set of
disjoint loops. By construction, loops are both self- and 
mutually-avoiding. Alternatively, a loop configuration may be viewed as a set
of {\it covered} edges such that each vertex is incident to either $0$
or $2$ covered edges.

\fig{A configuration of the $O(n)$ loop model on tetravalent maps,
when viewed on the dual quadrangulation, is built out of three types of
squares: (a) empty squares, weighted by $g$, (b) squares visited by a
loop going straight, weighted by $h_1$, and (c) squares visited by
a loop making a turn, weighted by $h_2$.}{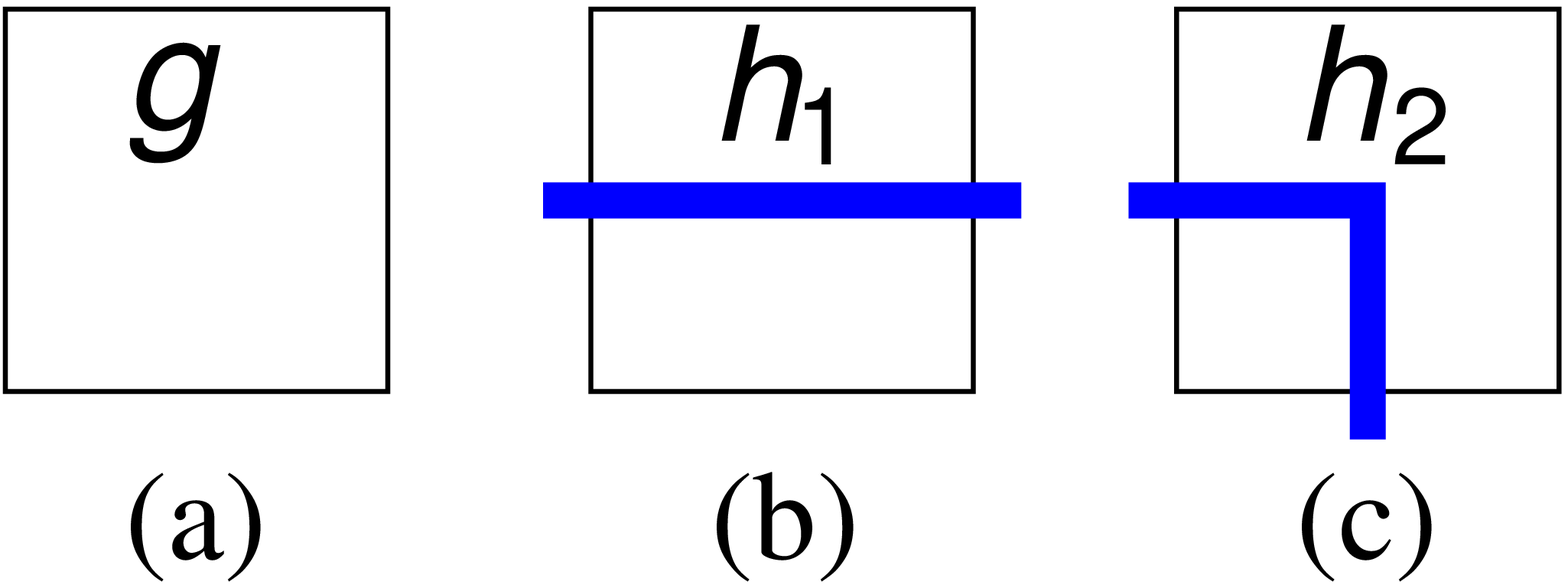}{10.cm}
\figlabel\squares

We prefer to work with the dual map which is a planar quadrangulation.
Then, the loops cross some edges of the quadrangulations. Up to a
rotation, the quadrangulation has three possible types of faces, as
displayed in Fig.~\squares: (a) empty squares not visited by a loop,
(b) squares visited by a loop going straight (i.e. crossing opposite
sides of the square) and (c) squares visited by a loop making a turn
(i.e. crossing consecutive sides of the square). The length of a loop,
i.e. the total number of faces that it visits, may be even or
odd. However, the number of faces of type (b) visited by a given loop is
necessarily even, resulting from the fact that a planar
quadrangulation is bipartite (other interesting consequences of the
bipartite nature of our maps will be seen below). A loop is said {\it rigid}
if it only visits faces of type (b).

Generally speaking, the $O(n)$ loop model consists in attaching a
non-local weight $n$ to each loop, in addition to some local weights.
Here, we consider an ``annealed'' model where the map and the loops are
drawn at random altogether: a configuration of the model is the data
of a planar tetravalent map endowed with a loop configuration. Local
weights are attached to the vertices of the tetravalent map, or
equivalently to the faces of the dual quadrangulation,
and naturally these weights depend on the
face type, see again Fig.~\squares: $g$ per square of type (a), $h_1$
per square of type (b), $h_2$ per square of type (c). The global
weight of a configuration is the product of all loop and face weights,
and the generating (``partition'') function of the model is the sum of
the global weights over all configurations. Here $n,g,h_1,h_2$ are
taken as non-negative real numbers. The model is said {\it well-defined}
when the generating function is finite
(this is the case for instance when $g + \max(n,1) (2h_1+4h_2) \leq
1/12$). It is then possible to normalize the configuration weights and
define a probability distribution. The {\it rigid loop model}
corresponds to taking $h_2=0$ so that loops are obliged to go straight
within each visited square.

\fig{A loop configuration on a quadrangulation with a boundary of length $2p=8$,
with $3$ loops, $9$ squares of type (a), $8$ squares of type (b) and $11$ squares of
type (c). Its weight is thus $n^3 g^9 h_1^8 h_2^{11}$. The associated gasket
consists of $5$ regular squares and $2$ holes with degrees $6$ and $10$.}{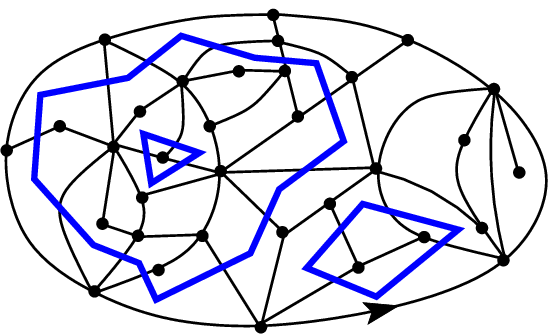}{10.cm}
\figlabel\loopconfig
As usual it is easier to work with rooted maps. Recall that a rooted
map has a distinguished oriented edge (the {\it root edge}). From now on,
the face on the right of the root edge will be called the
{\it external face} and we
define a {\it quadrangulation with a boundary of length $2p$} ($p \geq
1$) as a rooted bipartite planar map where the external face has
degree $2p$ and every other face has degree 4.  We shall now consider loop
configurations on the dual of a quadrangulation with a boundary (see Fig.~\loopconfig) and,
for simplicity, we assume that no loop visits the external face, to
which we therefore decide to attach a local weight $1$. Keeping the
same weights $n,g,h_1,h_2$ as above, we define $F_p^{\rm
loop}(n;g,h_1,h_2)$ as the generating function for the $O(n)$ loop
model on duals of quadrangulations with a boundary of length $2p$. As
particular cases, for $p=1$ we obtain (upon squeezing the bivalent external
face) rooted quadrangulations where the root edge is not crossed by a loop,
while for $p=2$ we obtain (directly) rooted quadrangulations
where the external face is not visited.

\subsec{The gasket decomposition}

The fundamental observation of this paper is that 
$F_p^{\rm loop}(n;g,h_1,h_2)$ is related the generating function $F_p$ of bipartite 
planar maps {\it without loops} with a boundary of length $2p$, defined as follows. 
Let us recall that a planar map is bipartite if and only if all its faces have even degree. 
We say that it has a boundary of length $2p$ if it is rooted  and the external face,
defined again as the face on the right of the root edge, 
has degree $2p$. To each internal face of degree $2k$, ($k \geq 1$), 
we attach a weight $g_k$, and the weight of a rooted map is the product of all 
its internal face weights. We then define $F_p$ as the sum of the weights of all 
bipartite maps with a boundary of length $2p$. 
Our main statement is that
\eqn\Fkloop{F_p^{\rm loop}(n;g,h_1,h_2)=F_p(g_1,g_2,\ldots)}
for the particular sequence $(g_k)_{k\geq 1}$ of
face weights satisfying the fixed point condition
\eqn\fixepoint{g_k= g
\delta_{k,2}+n \sum_{k'\geq 0} A_{k,k'}(h_1,h_2) F_{k'}(g_1,g_2,\ldots),\quad
k\geq 1\ ,}
with $A_{k,k'}(h_1,h_2)$ a polynomial in $h_1$ and $h_2$ whose
expression will be given in the next Section. For the rigid case, it reads
simply $A_{k,k'}(h_1,0)=h_1^{2k}\delta_{k,k'}$, leading to the simpler
fixed point condition
\eqn\fixepointrigid{g_k= g \delta_{k,2}+n h_1^{2k} F_{k}(g_1,g_2,\ldots)\ .}

\fig{The outer and inner contours of a loop.}{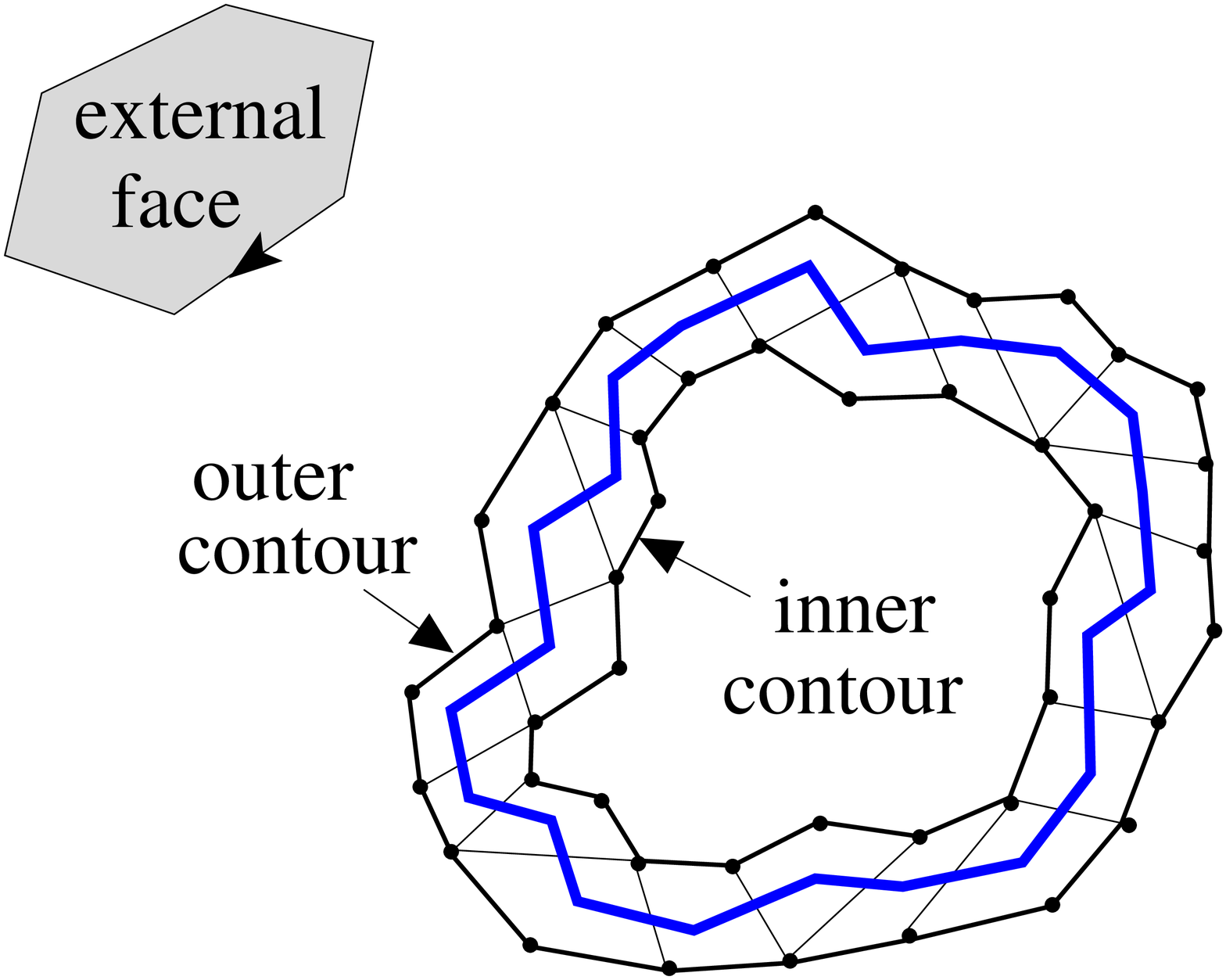}{7.cm}
\figlabel\contours
Let us now justify our fundamental observation, by considering a
quadrangulation with a boundary of length $2p$ endowed with a loop
configuration. For conciseness, we will often, in this Section, call
simply ``quadrangulation'' a quadrangulation with a boundary. Thanks to
the external face, the notions of exterior and interior of a loop are
well-defined. To each loop, we may associate its {\it outer}
(resp. {\it inner}) {\it contour} formed by the edges of the
quadrangulation which (i) belong to the exterior (resp. interior) of
the loop and (ii) are incident to squares visited by the loop (see
Fig.~\contours). Note that these contours are closed paths living on the
quadrangulation itself and therefore each of them has an even length. Note
also that the inner contour may be empty, i.e. may have length $0$ if
the loop encircles a single vertex.
\fig{The gasket is obtained by wiping out the content of
the outer contours of the outermost loops, creating holes.}{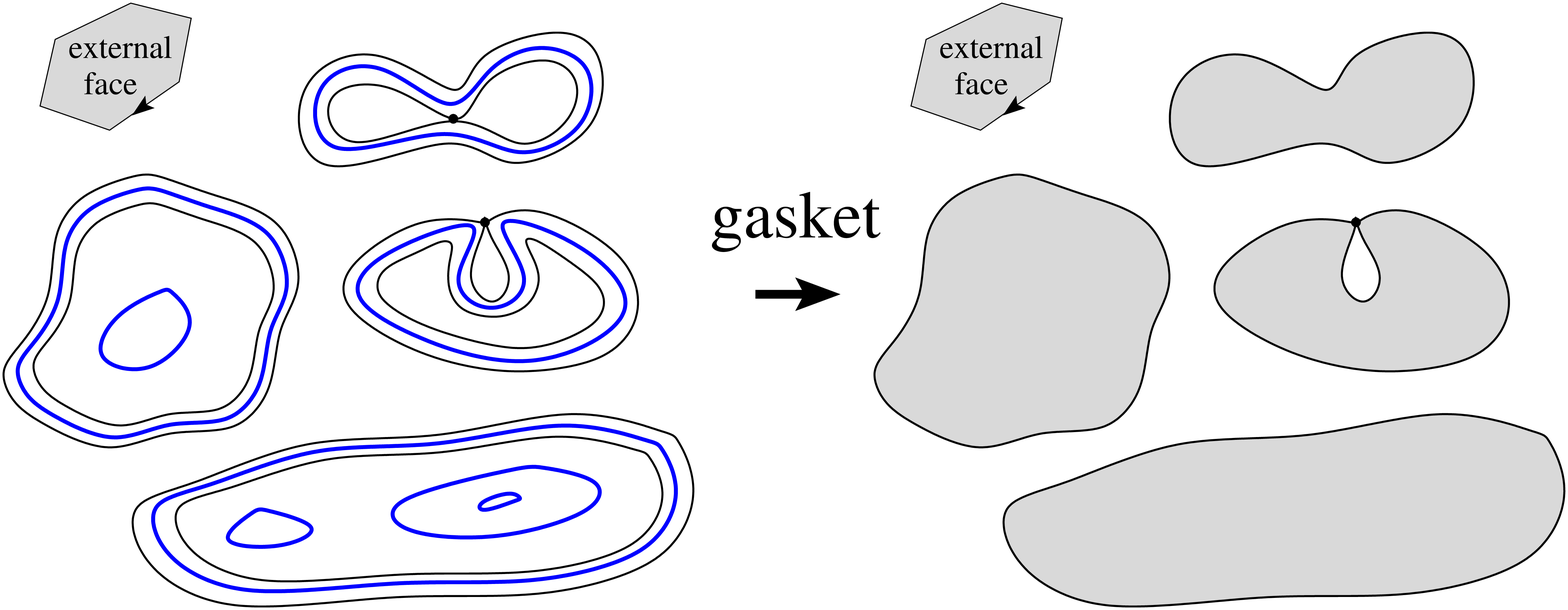}{12.cm}
\figlabel\gasket
Using a terminology similar to \LGM, we call {\it gasket}
the map formed by edges that are exterior to all the loops, see
Fig.~\gasket\ for an illustration.  It is a rooted bipartite planar map
containing in particular all the edges incident to the external face,
which therefore remains of degree $2p$.  Its other faces are of two
types: {\it regular faces} which are the squares of the original
quadrangulation that lay outside of every loop, and {\it holes}
delimited by the outer contours of the former outermost
loops.

Clearly, the transformation that goes from the original
quadrangulation with loops to the gasket is not
reversible. To make it reversible, we must
keep track of the former content of the holes. In this respect, a
first remark is that the interior of a loop may itself be
viewed as a quadrangulation with a boundary endowed with a loop
configuration, where the boundary is nothing but the inner contour of
the loop at hand (when the inner contour has length $0$, we have the
so-called ``vertex-map'' reduced to one vertex and one face, with no edge).
In order not to lose any information, we must also
consider the squares covered by the loop itself, which form a {\it
ring} lying in-between the inner and outer contours. As shown in
Fig.~\contours, this ring is a cyclic sequence of squares glued
together. To summarize, the content of a hole of degree $2k$ ($k\geq 1$)
is described by a pair formed by a ring with outer length $2k$ and
inner length $2k'$ (for some $k'\geq 0$), and an {\it internal}
quadrangulation with a boundary of length $2k'$ (equal to the vertex-map
if $k'=0$) endowed with a loop configuration.
\fig{Illustration of the rooting procedure. We draw the leftmost shortest
path (dashed line) in the gasket starting with the map root edge and
ending with an edge of the outer contour of the loop at hand, which we pick
as root for the ring.
We then select an edge of the inner contour by the rules displayed at the
bottom left, which we pick as root for the internal
quadrangulation.}{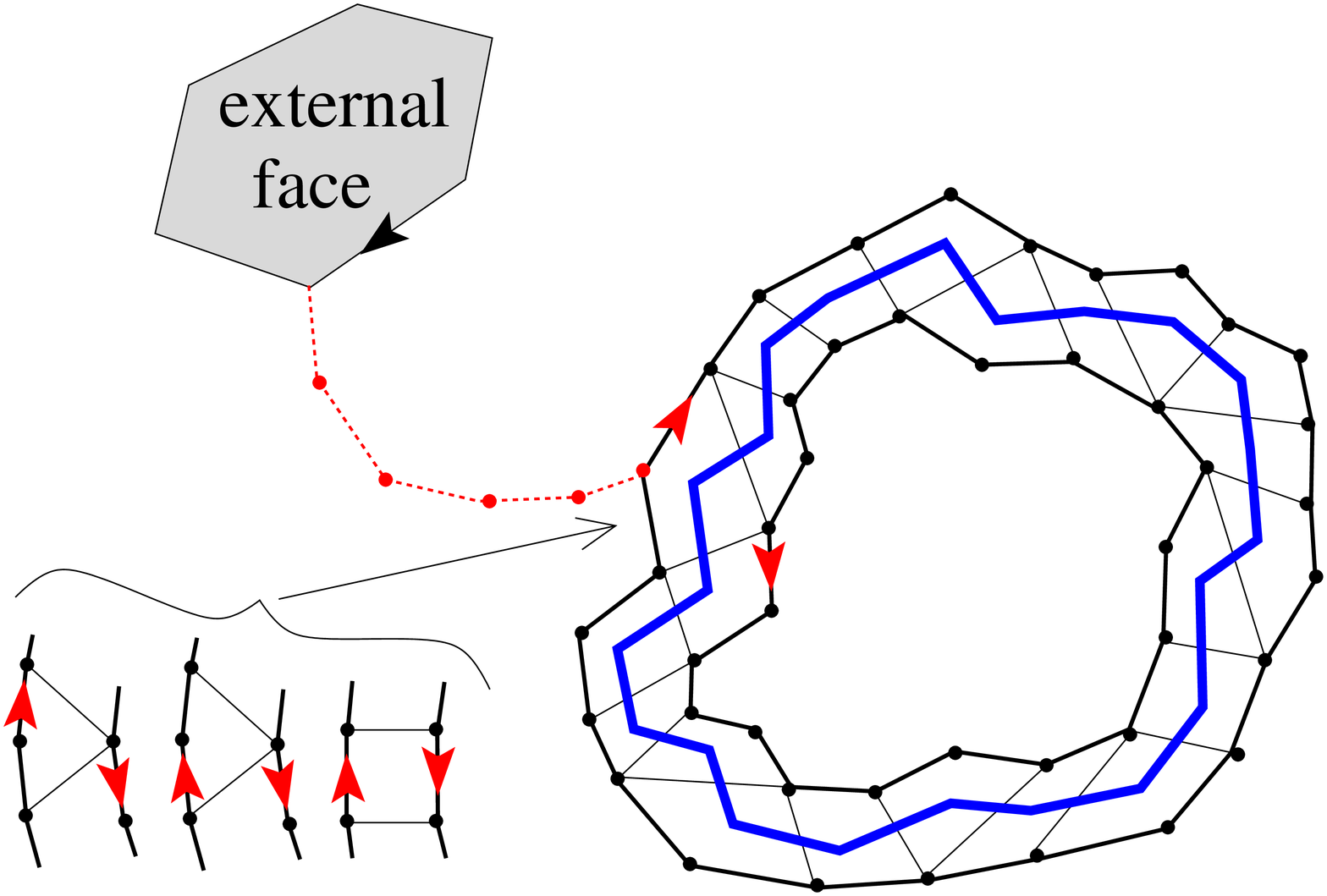}{8.5cm}
\figlabel\rooting
Two important remarks are in order. First, to ensure bijectivity, both
the ring and the internal quadrangulation must be rooted, respectively
on its outer contour and on its boundary. The positions of their root
edges is inherited from the rooting of the original quadrangulation
via a somewhat irrelevant yet well-defined procedure: for instance we
may consider the leftmost shortest path that stays within the
gasket, starts with the root edge and ends with an edge of the outer
contour, that edge being selected as the ring root; then, using only
the data of the ring, we may easily select an edge of the inner
contour which serves as root for the internal quadrangulation (see
Fig.~\rooting). Second, an essential assumption is that the original map, the
gasket and the internal quadrangulations are all possibly
separable (i.e. may contain separating vertices whose removal
disconnects them). In particular, a separating vertex incident to a
hole in the gasket appears whenever a multiple point was present along
the outer contour of the associated loop on the original
quadrangulation.  Similarly, a multiple point of the inner contour of
a loop results into a separating vertex on the boundary of the
associated internal quadrangulation. In contrast, all the vertices of
a ring are considered as distinct (since the information about
contacts is recorded in the gasket and internal quadrangulation).
At this stage, it should be clear that the decomposition is
reversible: given a ring and a quadrangulation with a boundary with
compatible lengths, there is a well-defined procedure to join them and
fill a hole of the gasket.

\fig{A face of the gasket is either a regular square (weighted by $g$)
or a hole of even degree, say $2k$. The content of this hole is made
of a ring with outer length $2k$ and inner length $2k'$ for some $k'\geq 0$
(weighted by $n A_{k,k'}(h_1,h_2)$), and of an internal quadrangulation
with a boundary of length $2k'$ (weighted by $F_{k'}^{\rm loop}(n;g,h_1,h_2)=
F_{k'}$).}{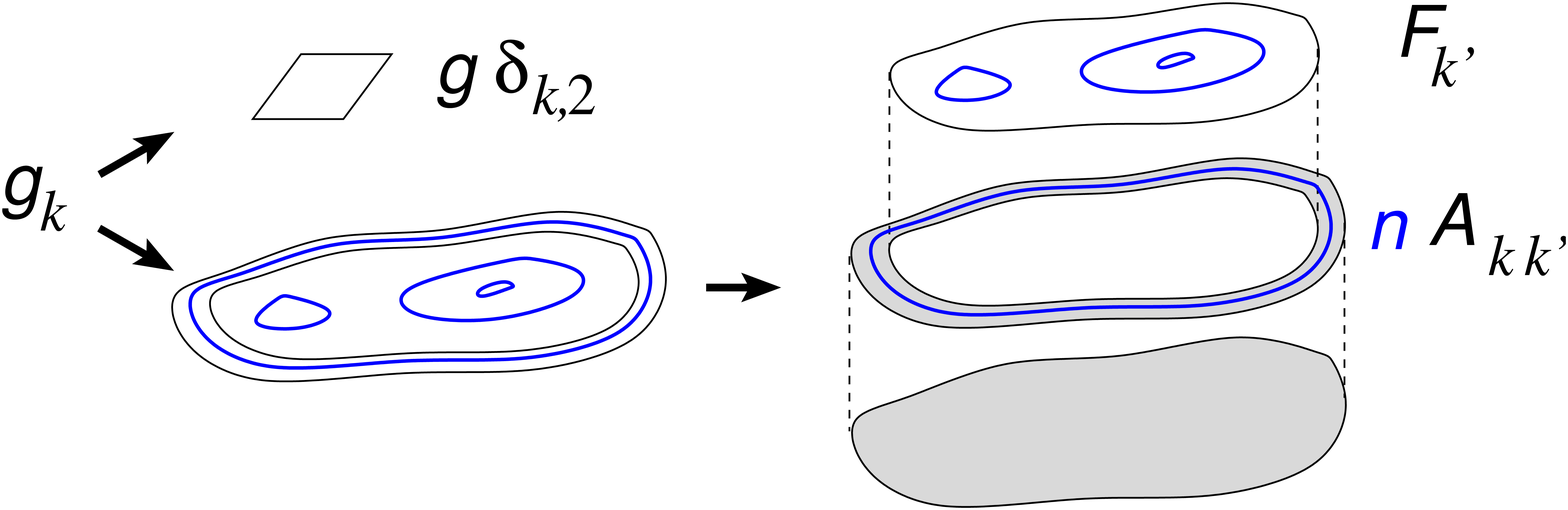}{12.cm}
\figlabel\holefill

Globally, we have a one-to-one correspondence between, on the one
hand, quadrangulations with a boundary endowed with a loop
configuration and, on the other hand, gaskets endowed with
hole contents.  Translating this correspondence in the language of
generating functions, we deduce that the wanted quantity $F_p^{\rm
loop}(n;g,h_1,h_2)$ is equal to the generating function $F_p$ for
gaskets, where each regular face still receives a weight $g$
and each hole of degree $2k$ receives a weight $n \sum
A_{k,k'}(h_1,h_2)F_{k'}^{\rm loop}(n;g,h_1,h_2)$ (see Fig.~\holefill).  In this hole
weight, $n$ accounts for the associated loop, $F_{k'}^{\rm loop}(n;g,h_1,h_2)$ for the
internal quadrangulation, while $A_{k,k'}(h_1,h_2)$ is the generating
function for rings with sides of lengths $2k$ and $2k'$.  This
establishes Eqs.~\Fkloop\ and \fixepoint.

\subsec{The ring generating function}
Let us now discuss the precise form of the ring generating function
$A_{k,k'}(h_1,h_2)$.
We have the explicit expression
\eqn\akkprime{A_{k,k'}(h_1,h_2)=\sum_{j=0}^{\min(k,k')}
{2k\over k+k'}{(k+k')!\over (2j)!(k-j)!(k'-j)!} h_1^{2j}
h_2^{k+k'-2j}\ ,}
which follows from a simple counting argument: the term with
index $j$ in the sum corresponds to rings with $2j$ squares of
type (b), each weighted by $h_1$. Such rings have $k-j$ (resp.\ $k'-j$)
squares of type (c) with their two unvisited sides
along the outer (resp.\ inner)
contour of the ring, all having a weight $h_2$.
There are ${k+k'\choose 2j,k-j,k'-j}$ possible sequences of
such three types of squares. However, this number has to be
corrected because a ring has a distinguished edge (among $2k$)
on its outer contour rather than a distinguished square (among $k+k'$)
to start the sequence. This explains the corrective factor $2k/(k+k')$.

\fig{The ring transfer matrix $M(z)$. An edge of the outer contour (thick
edges) is in state $1$ if it follows clockwise a vertex adjacent
to a vertex of the inner contour, and in state $2$ otherwise. The dashed lines
indicate how we distribute the weights in the transfer matrix. Each edge
in state $2$ is followed by an edge in state $1$ and receives the weight
$h_2$ of its incident square of type (c). Each edge in state $1$ first
receives a weight $1/(1-h_2 z^{-1})$ accounting for a sequence of
squares of type (c) as shown here in grey, then a further weight $h_1 z^{-1/2}$
if and only if it is followed by another edge in state $1$.
 }{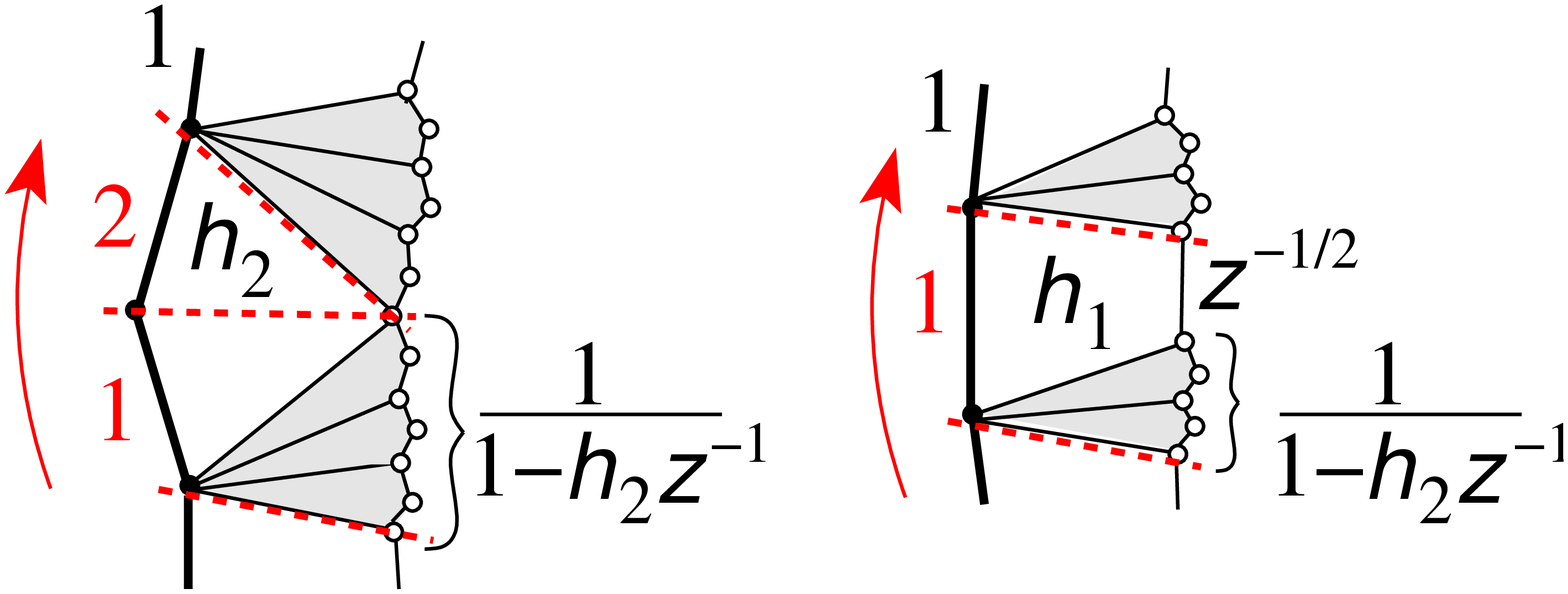}{11.cm}
\figlabel\transmat
We have the alternative expression, which will prove useful in the following:
\eqn\akkx{\sum_{k'\geq 0} A_{k,k'}(h_1,h_2)
z^{-k'}\!=\!(\lambda_+(z))^k\!+\!(\lambda_-(z))^k, \quad
\lambda_\pm(z)\!=z\left({h_1 \pm \sqrt{h_1^2\!-\!4 h_2^2\!+\!4 h_2 z}
\over 2 (z-h_2)}\right)^2\ }
(note that the sum converges for $|z|>h_2$). This expression naturally
follows from a transfer matrix argument. As seen in Fig.~\transmat, we
may distinguish two ``states'' for the edges of the outer contour depending
on whether they follow clockwise a vertex adjacent to a vertex of
the inner contour (state 1) or not (state 2). We may then write
\eqn\akktrans{\sum_{k'\geq 0} A_{k,k'}(h_1,h_2)
z^{-k'}\!=\!\tr M(z)^{2k}, \qquad
M(z)\!=\pmatrix{{\displaystyle{h_1 z^{-1/2} \over 1-h_2 z^{-1}}}&
\displaystyle{{1 \over 1-h_2 z^{-1}}}
\cr & \cr h_2 & 0 \cr}\ ,}
where $M(z)$ is the transfer matrix describing the transition between
two successive edge states. This establishes \akkx\ where $\lambda_\pm(z)$
are simply the eigenvalues of $M^2(z)$, solutions of:
\eqn\detakk{\left( h_2 (\lambda+z) - \lambda z \right)^2 - h_1^2 \lambda
z =0\ .}
Interestingly, $\lambda$ and $z$ play a symmetric role in this equation.
It follows that
$\lambda(\lambda(z))=z$ provided we pick the correct determination of $\lambda$.
More precisely, it may be checked that $z\mapsto \lambda_+(z)$ is an involution on
$(h_2,\infty)$ with a unique fixed point:
\eqn\xstar{z^*=h_1+2h_2=\lambda_+(z^*)\ .}

Finally, we observe that for rigid loops, there is a unique ring
with outer length $2k$, and it has the same inner length. This yields
immediately $A_{k,k'}(h_1,0)=h_1^{2k}\delta_{k,k'}$ as announced in
the previous Section. This is compatible with \akkprime\ for
$h_2=0$, and with \akkx\ upon noting that
$\lambda_+(z)=h_1^2/z$ and $\lambda_-(z)=0$.

\newsec{Bipartite maps with arbitrarily large faces: possible critical
behaviors}

In this Section, we step away from the $O(n)$ loop model and discuss
a few properties of the generating functions of bipartite maps.

\subsec{Reminders}

As already mentioned, a planar map is bipartite if and only if all its faces have even degree. 
To each face of degree $2k$, ($k \geq 1$), we attach a weight $g_k$, here
considered as an arbitrary non-negative real number. Furthermore, we will sometimes
find convenient to also attach a non-negative
real weight $u$ per vertex of the map, irrespectively of its degree. The
unnormalized weight of a map is then the product of all its face and vertex
weights. Note that,
from Euler's relation, we could set $u=1$ without loss of generality upon
redefining $g_k \to g_k u^{k-1}$.

It is well-known that the simplest map enumeration formulas are for
{\it pointed rooted maps}, i.e. maps with a distinguished vertex (pointed)
and a distinguished oriented edge (rooted). The generating function
(i.e. the sum over all maps of the unnormalized weights above) for pointed
rooted bipartite planar maps is $2 u (R(u)-u)$ where $R(u)$ is the smallest
non-negative (possibly infinite) solution of the equation \MOB 
\eqn\eqforR{R(u)=u+\sum_{k\geq 1} g_k {2k-1 \choose k} R(u)^k\ .}
Here, we choose to display explicitly the dependence in $u$
but to hide that in the $g_k$'s. It is easily seen that $R(u)$ is
an increasing function of $u$ with $R(0)=0$.

Other more involved generating functions can be expressed in terms of $R(u)$.
For instance, the generating function $F_k$ for maps with a boundary of length $2k$, 
as defined in the previous Section (with $u=1$), reads
\HANKEL
\eqn\eqforFk{F_k={2k \choose k} \int_0^1 R(u)^k du\ .}
Consistently, we set $F_0=1$. 
Furthermore,
since rooted maps are obviously in bijection with maps with a boundary of
length $2$ and not reduced to a single edge, we may interpret $F_1-1$
as the generating function for rooted maps. Upon integrating \eqforR\ and
noting that ${2k \choose k}=2{2k-1 \choose k}$, we immediately deduce
\eqn\Fone{F_1-1= \sum_{k\geq 1} g_k F_k\, }
which amounts to decomposing the generating function of rooted maps (l.h.s)
according to the root degree of the maps (r.h.s).

\subsec{Subcritical and generic critical ensembles}

We say that the sequence $(g_k)_{k\geq 1}$ of face weights is {\it admissible}
if $R(1)$ is finite. In this case, taking $u=1$ and dividing the unnormalized
weight of a pointed rooted map by $2(R(1)-1)$, we obtain its probability
in the {\it Boltzmann ensemble} of pointed rooted maps associated with the
sequence $(g_k)_{k\geq 1}$.
For example, random quadrangulations are obtained by choosing the
sequence $g_k=g\, \delta_{k,2}$ ($g>0$), which is admissible for
$g\leq 1/12$.

A number of properties of the Boltzmann ensemble are encoded in the expansion
of $R(u)$ around $u=1$. This expansion is obtained upon inverting the expansion
around $R=R(1)$ of the inverse function $u(R)$ which, from \eqforR, reads
explicitly:
\eqn\eqforu{u(R)= R-\varphi(R), \quad \varphi(R)=\sum_{k\geq 1}
g_k {2k-1 \choose k} R^k\ .}
Here the function $\varphi(R)$ is defined through its series expansion at $R=0$,
whose coefficients are all non-negative. Its radius of convergence is
$R_c=1/(4 \lim \sup_{k\to \infty} (g_k)^{1/k})$ since 
${2k \choose k}\sim 4^k/\sqrt{\pi k}$ at large $k$.
If the sequence is admissible, $R_c$
is necessarily non-zero (possibly infinite), with $R(1)\leq R_c$, and
it is easily seen that $u'(R(1))\geq 0$, i.e. $\varphi'(R(1))\leq 1$.
The sequence is said {\it critical} if $\varphi'(R(1))= 1$
(hence $u'(R(1))=0$ and $R'(u)\to \infty$ as $u\to 1$) and
{\it subcritical} otherwise
(hence $R'(1)$ is finite).  This change of behavior is visible in the
large $k$ asymptotics of $F_k$. Indeed, the integral \eqforFk\ is
asymptotically dominated by the vicinity of $u=1$ and its behavior
may be computed via Laplace's method \Dieudonne. 
For a subcritical sequence, we
have at large $k$:
\eqn\asympFksub{F_k \sim {R(1)(1-\varphi'(R(1))\over \sqrt{\pi}}{(4 R(1))^k
\over k^{3/2}}\ .}
This form of the asymptotics is observed for instance when
all $g_k$'s are zero, in which case $\varphi(R)=0$, $R(u)=u$, and $F_k=
{2k \choose k}/(k+1)$ is the $k$-th Catalan number.

For a critical sequence, the asymptotic behavior depends on whether
the (negative) quantity $u''(R(1))=-\varphi''(R(1))$ is finite or not.
For a {\it generic critical sequence}, this quantity is finite and we
have that
\eqn\asympFkcritgen{F_k \sim {R(1)^2\varphi''(R(1))\over \sqrt{\pi}}
{(4 R(1))^k
\over k^{5/2}}\ .}
This form of the asymptotics is observed for instance for
critical random quadrangulations obtained by choosing the sequence
$g_k=(1/12) \delta_{k,2}$,
in which case $\varphi(R)=R^2/4$, $R(1)=2$, $\varphi''(R(1))=1/2$, so that
\asympFkcritgen\ is consistent
with the exact formula $F_k= 2^{k+1} {(2k)!\over k!(k+2)!}$.

In order to escape from this generic critical behavior, we must have
$\varphi''(R(1))$ infinite, which implies that $R(1)$ must coincide with
the radius of convergence $R_c$ of $\varphi$. Such non-generic
critical behaviors are analyzed in the next section. Before
proceeding, let us mention the following simple lemma:

\medskip
\noindent $\bullet$ {\bf Monotonicity property:}
if $(g_k)_{k \geq 1}$ and $(\tilde{g}_k)_{k \geq 1}$ are
two sequences such that $g_k \leq \tilde{g}_k$ for all $k$, the
inequality being strict for at least one $k$, and if $(\tilde{g}_k)_{k
\geq 1}$ is admissible, then $(g_k)_{k \geq 1}$ is subcritical.
\medskip

Indeed, denoting by $\tilde{R}(\cdot)$ and $\tilde{\varphi}(\cdot)$
the functions associated with the sequence $(\tilde{g}_k)_{k \geq 1}$,
we have $R(1) < \tilde{R}(1)$ since $2R(1)-1$, being a map generating 
function, is a strictly increasing function of $g_k $ for each $k$. 
Then, $\varphi'(R(1)) <
\tilde{\varphi}'(\tilde{R}(1)) \leq 1$ hence, by definition, $(g_k)_{k
\geq 1}$ is subcritical.

\subsec{Non-generic critical ensembles}

In order to get a non-generic critical behavior, we must ensure
simultaneously the two conditions $\varphi'(R(1))=1$ (criticality)
and $\varphi''(R(1))=\infty$ (non-genericity).
This turns out to highly constrain the possible form of the sequence
$(g_k)_{k\geq 1}$. As shown by Le Gall and Miermont \LGM, a natural
candidate for such a sequence is:
\eqn\gknongen{g_k=c \left({1\over 4 R_c}\right)^{k-1} g_k^\circ}
where $(g_k^\circ)_{k\geq 1}$ is an arbitrary sequence of reference, such that
\eqn\gkcirc{g_k^\circ \sim k^{-a} \quad \hbox{for}\ k\to\infty\ \quad
\hbox{with}\  {3\over 2}<a<{5\over 2}\ .}
Note that $R_c$ is indeed the radius of convergence of $\varphi(R)$ since
we have $\varphi(R)= c R f_\circ(R/(4 R_c))$, where
\eqn\fcirc{f_\circ(x)=\sum_{k\geq 1}
{2k-1 \choose k} g_k^\circ\, x^{k-1}}
is an analytic function with radius of convergence $1/4$.
Since $a$ lies in the range $]3/2,5/2[$, we have $\varphi'(R_c)<\infty$
and $\varphi''(R_c)=\infty$. The non-genericity condition therefore
reduces to demanding that
\eqn\RoneRc{R(1)=R_c\ ,}
or equivalently $1=R_c-\varphi(R_c)$. Together with the criticality condition
$\varphi'(R_c)=1$, this fixes the values of $c$ and $R_c$ as
\eqn\valcbeta{c={4\over 4 f_\circ(1/4)+f_\circ'(1/4)}, \quad
R_c=1+{4 f_\circ(1/4)\over f_\circ'(1/4)}\ .}
At these values, by standard transfer theorems \FS, $u(R)$ has a singular expansion around $R=R_c$
given by
\eqn\expansing{u(R)=1-{2 c\, R_c\, \Gamma(1/2-a) \over
\sqrt{\pi}} \left(1-{R\over R_c}\right)^{a-1/2}+O\left(1-{R \over R_c}
\right)^2\ .}
Note that $\Gamma(1/2-a)>0$ since $3/2<a<5/2$.
Inverting this expansion, we see that the generating function $R(u)$
behaves, when we fix the face weights $g_{k}$ as above and
let the vertex weight $u$ tend to $1$ from below (i.e. we use
the parameter $u$ to control the approach to the critical point), as
\eqn\Ruexp{R(u)=R_c(1-\kappa\, (1-u)^{2/(2a-1)})+O(1-u),
\quad \kappa=\left({\sqrt{\pi}\over 2 c\, R_c\, \Gamma(1/2-a)}
\right)^{2/(2a-1)}\ .}
Again, by transfer theorems, this implies that the coefficient
of $u^N$ in $R(u)$, hence the probability of having $N$ vertices
in the Boltzmann ensemble of pointed rooted maps, decays as
$N^{-(2a+1)/(2a-1)}$ as $N\to \infty$.

Returning to the asymptotics of $F_k$, we find by Laplace's method
that, for large $k$:
\eqn\Laplace{\int_0^1 R(u)^k du\sim {2 c R_c\, (a-1/2)\Gamma(1/2-a)\Gamma(a-1/2)
\over \sqrt{\pi}}{R_c^{k}\over k^{a-1/2}}\,}
and, using ${2k \choose k}\sim 4^k/\sqrt{\pi k}$, 
$(a-1/2)\Gamma(1/2-a)\Gamma(a-1/2)=\pi/\sin\pi(a-3/2)$ and $R_c=R(1)$,
\eqn\asympFkcritnongen{F_k\sim
{2 c \, R(1) \over \sin \pi (a-3/2)} {(4 R(1))^k\over k^a}\ .}
It is instructive to compare this expression with the ``input'',
namely the form \gknongen-\gkcirc\ for non-generic critical
face weights, which implies the asymptotics
\eqn\gkasymp{g_k\sim 4 c\, R(1)\, {(4 R(1))^{-k}\over k^a}\ .}
We note a striking similarity: the prefactor is simply divided by
$2 \sin \pi (a-3/2)$ while the exponential factor is inverted.
This similarity is not merely anecdotal but will have important 
consequences for the $O(n)$ loop model (see Section 4).

As a concluding remark, note that the condition $R(1)=R_c$ for non-generic
criticality implies that the face degree distribution has a fat tail.
As seen from the derivation \MOB\ of \eqforR, 
the probability that the external face in a rooted pointed 
map has degree $k$ is
\eqn\prprob{{g_k {2k-1 \choose k} R(1)^k \over R(1)-1 }} 
which decays exponentially as $(R(1)/R_c)^k$ when $R(1)<R_c$, but
only as a power law $k^{-a-1/2}$ at a non-generic critical point.
Similarly, from \Fone, the probability that the external face in a rooted 
(but unpointed) map has degree $k$ is
\eqn\prprob{{g_k F_k \over F_1-1 }}               
which also decays exponentially as $(R(1)/R_c)^k$ when $R(1)<R_c$ and
as a power law, now $k^{-2a}$, at a non-generic critical point.
Note finally that the condition $R(1)=R_c$ is necessary but not
sufficient for non-generic criticality as one may construct examples
of subcritical and generic critical sequences such that $R(1)=R_c$.
Such situations will however not appear in the context of the $O(n)$
model with non-negative weights.

\subsec{The resolvent}

For the purposes of Section 6, it is useful to gather facts about the 
so-called {\it resolvent} (this terminology being
borrowed from the matrix integral formalism), defined as:
\eqn\resolvdef{W(\xi) = \sum_{k \geq 0} {F_k \over \xi^{2k+1}}.}
Using \eqforFk\ and performing the change of variable $u \to
R$, we obtain
\eqn\resolvu{W(\xi) = {1 \over \xi} \int_0^{R(1)} {u'(R) dR \over
 \sqrt{1-4R/\xi^2}}.}
From this expression, it is seen that $W(\xi)$ is analytic in the
complex plane minus the segment $[-\gamma,\gamma]$, with
$\gamma=2\sqrt{R(1)}$. Along this segment,
it has a discontinuity encoded into the so-called {\it spectral density}
\eqn\rhodef{\rho(\xi)={W(\xi-{\rm i}0) - W(\xi+{\rm i}0) \over
  2{\rm i}\pi} = {1 \over \pi} \int_{\xi^2/4}^{R(1)} {u'(R) dR \over
 \sqrt{4R-\xi^2}}\ , \quad \xi \in [-\gamma,\gamma].}
Since $u'(R)\geq 0$ for $R \leq R(1)$, $\rho(\xi)$ is nonnegative on
$[-\gamma,\gamma]$. Note that, by applying the Cauchy formula
around the cut, we may recover $W(\xi)$ from $\rho(\xi)$ via
\eqn\Wrho{W(\xi)=\int_{-\gamma}^{\gamma} {\rho(\xi')
d\xi' \over \xi - \xi'}\ ,\quad \xi \notin [-\gamma,\gamma]}
and in particular, since $W(\xi) \sim 1/\xi$ for $\xi \to
\infty$, we have the normalization $\int_{-\gamma}^{\gamma} \rho(\xi) d\xi=1$.
The asymptotics \asympFksub, \asympFkcritgen\ and \asympFkcritnongen\ translate 
into the respective
singularities of $\rho(\xi)$ at $\xi=\pm \gamma=\pm 2\sqrt{R(1)}$:
\eqn\rhosing{\eqalign{\rho(\xi) &\sim {1 - \varphi'(R(1)) \over 2
\pi} \left(4 R(1) -\xi^2\right)^{1/2} \quad \hbox{subcritical} \cr \rho(\xi)
&\sim {\varphi''(R(1)) \over 12 \pi} \left(4 R(1) -\xi^2\right)^{3/2} \quad
\hbox{generic critical}\cr
\rho(\xi) & \sim {4^{1-a} R(1)^{3/2-a} c \over \Gamma(a)\, \sin \pi
(a-3/2)}(4 R(1) - \xi^2)^{a-1} \quad \hbox{non-generic critical} \ .\cr}}

\newsec{Asymptotic self-consistency}

In this Section, we return to our $O(n)$ loop model and make a first analysis of the fixed point
condition \fixepoint\ via its asymptotic consequences. The fixed point 
condition is an equation for the unknown sequence $(g_k)_{k\geq 1}$ 
that depends on the physical parameters
$n$, $g$, $h_1$ and $h_2$. By definition, the model is well-defined when
a finite solution exists (i.e. $F_p^{\rm loop}$ is finite).
Since it involves the generating
functions $(F_k)_{k\geq 0}$, such a finite solution $(g_k)_{k\geq 1}$
is necessarily admissible in the sense of Section 2. The purpose of
this Section is to classify the possible nature (subcritical or critical,
generic or not) of the solution.

\subsec{Rigid case}

For simplicity we first concentrate on the rigid case. In this
Section, $(g_k)_{k \geq 1}$ denotes a sequence satisfying the fixed
point condition \fixepointrigid. It immediately implies the
asymptotic condition
\eqn\gkasymrig{g_k \sim n h_1^{2k} F_k \qquad (k \to \infty).}
In Section 3, we have classified the possible asymptotic behaviors of
$F_k$ depending on the nature of the weight sequence $(g_k)_{k\geq
1}$: the equations \asympFksub, \asympFkcritgen\ and
\asympFkcritnongen\ describe respectively the subcritical, generic
critical and non-generic critical cases. Here we assume that, in the
non-generic critical case, the weight sequence is of the form 
\gknongen-\gkcirc. It is shown in Section 6 that no other more exotic 
non-generic behavior exists.

We now remark that Eq.~\eqforFk\ implies that $(F_k)^{1/k} \to 4 R(1)$ for large $k$, 
regardless of the nature of the weight sequence.
By comparing with the
general relation $\lim \sup_{k\to \infty} (g_k)^{1/k} = 1/(4 R_c)$, where $R_c$ is the
radius of convergence of the function $\varphi(R)$ defined in
Eq.\eqforu, Eq.\gkasymrig\ implies the relation
\eqn\rigidcond{16 h_1^2 R_c R(1) = 1\ .}
For later use, we introduce the parameter 
\eqn\taudefrigid{\tau= 4 h_1 R(1)\ .}
Since $R(1) \leq R_c$, we have by \rigidcond\ that $\tau\leq 1$ and
the condition $\tau=1$ amounts to $R(1)=R_c$. 
We now prove the following:

\medskip
\noindent $\bullet$ {\bf Non-genericity criterion}: 
$R(1)=R_c$ (i.e.\ $\tau=1$) 
if and only if the sequence $(g_k)_{k\geq
1}$ is critical and non-generic. 

\medskip
The first implication (``if'') was
shown in Section 3 without recourse to the fixed point
condition (see Eq.~\RoneRc). 
Conversely, assume that the sequence $(g_k)_{k\geq 1}$ is
either subcritical or generic critical. Its asymptotic behavior is
directly deduced from that of $F_k$ via Eq.\gkasymrig. In the
subcritical case, due to the $k^{-3/2}$ factor in Eq.\asympFksub,
$\varphi'(R_c)$ is infinite while, by definition of subcriticality, we
have $\varphi'(R(1))<1$, hence $R(1) \neq R_c$. In the generic
critical case, due to the $k^{-5/2}$ factor in Eq.\asympFkcritgen,
$\varphi''(R_c)$ is infinite while, by definition of genericity, we
have $\varphi''(R(1))<\infty$, hence $R(1) \neq R_c$. This completes
the proof (note that when we do not impose the
fixed point condition, we can easily construct examples of subcritical and
generic critical sequences such that $R(1)=R_c$).
In view of the discussion at the end of Section 3.3, we deduce that 
the face degree
distribution of the gasket, i.e. the outer loop length distribution 
in the $O(n)$ model, has a fat tail iff the model is at a non-generic critical
point.

If we now assume that the sequence $(g_k)_{k\geq 1}$ is critical and
non-generic, and takes the general form \gknongen\ for a suitable
sequence $(g_k^\circ)_{k\geq 1}$ satisfying \gkcirc, then, in addition
to the condition $\tau=1$, we see by comparing the prefactors of 
\asympFkcritnongen\ and \gkasymp\ that 
\eqn\valn{n=2 \sin \pi (a-3/2)\ .}
In particular, $n$ must be in the range
$]0,2[$ for non-generic criticality to be possible.

To summarize, we have shown that the fixed point condition
\fixepointrigid\ is compatible with the model being
either subcritical, generic critical, with $\tau <1$ or, more interestingly,
non-generic critical with $\tau=1$ and $a=2 \pm b$, $\pi b = \arccos(n/2)$. 

So far, we did not tell which behavior corresponds to a given value of
the ``physical'' parameters $n$, $g$ and $h_1$. Since the generating
function $F_p^{\rm loop}(n;g,h_1,0)$ is an increasing function of
its parameters, there exists a decreasing function $h_c(n;g)$ such
that the model is well-defined for $h_1<h_c(n;g)$ and ill-defined for
$h_1>h_c(n;g)$. In particular, $h_c(n;g)>0$ if $g < 1/12$. By the
monotonicity property of Section 3.2, we see that the model is
subcritical whenever $h_1<h_c(n;g)$. In particular, any critical point 
necessarily lies on the line $h_1=h_c(n;g)$. An exact expression for
$h_c(n,g)$ will be given in Section 6 for $0<n<2$. There we find that the 
model is critical all along the line $h_1=h_c(n;g)$. Furthermore, along
this line, there exists a 
$g^*$ such that (i) for $g^* < g \leq 1/12$ the model is generic critical, 
(ii) for $0 \leq g<g^*$ the model is non-generic critical with $a=2-b$
({\it dense} model) and (iii) at $g=g^*$ the model is non-generic
critical with $a=2+b$ ({\it dilute} model). The corresponding phase diagram
is displayed in Fig.12 below.

\subsec{Non-rigid case}
Let us now consider the general case of arbitrary non-negative values
of $h_1$ and $h_2$ and analyze the consequences of the fixed point
condition \fixepoint\ on the asymptotics of $g_k$. To treat the subcritical,
generic critical and non-generic critical cases simultaneously, we write the
asymptotics of $F_k$ in the form
\eqn\Fkall{F_k\sim \chi {(z_c)^{-k}\over k^a},\quad z_c={1\over 4R(1)}\ ,}
with now $3/2\leq a\leq 5/2$ and where $\chi$ may be read off
Eqs.\asympFksub, \asympFkcritgen\ and \asympFkcritnongen\ respectively.
Here again, we assume that, in the non-generic critical case, the weight 
sequence is of the form \gknongen-\gkcirc. 

Let us introduce the generating function
\eqn\genF{F(z)=\sum_{k\geq 0} F_k\, z^k\ ,}
related to the resolvent via $W(\xi)=F(1/\xi^2)/\xi$. By transfer
theorems, the
function $F(z)$ is singular when $z\to z_c$ with a singular part given by
\eqn\singF{F(z)\vert_{\rm sing}\sim \chi\, \Gamma(1-a)\, \left(1-
{z\over z_c}\right)^{a-1}}
where we temporarily exclude the case $a=2$.
\fig{Deformation of the contour of integration in Eq.~(4.8): starting from
a circle with radius between $h_2$ and $z_c$, we deform it by
encompassing the cut of the function $F(z)$, as shown on the right,
so that the dominant contribution for $k\to \infty$ is given,
after the change of variable $z\to \delta$ of Eq.~(4.9), by integrating
the leading order of the integrand over the Hankel contour
shown below. }{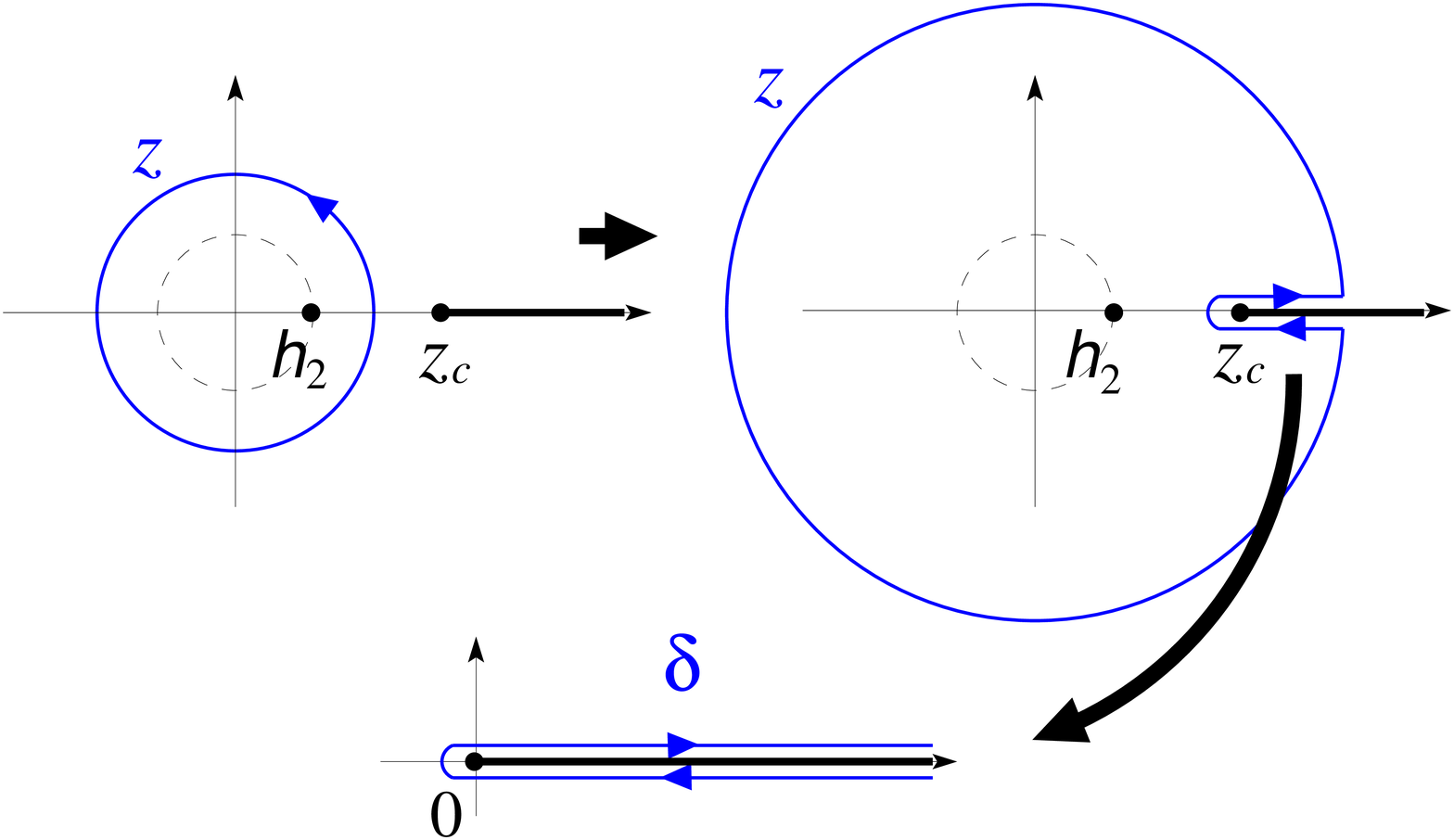}{10.cm}
\figlabel\intcont
\noindent To estimate the r.h.s of \fixepoint, we use Eq.\akkx\ to write
\eqn\contint{\sum_{k'\geq 0} A_{k,k'}(h_1,h_2) F_{k'}
= \oint {dz\over 2 {\rm i} \pi z} F(z)\, \left(\lambda_+(z)^k+\lambda_-(z)^k\right)}
where me may choose as integration contour the circle $|z|=z_0$
for any $h_2<z_0<z_c$ (note that, since $A_{k,k'}(h_1,h_2)\geq A_{k,k'}(0,h_2)=
2 {k+k'-1 \choose k-1} h_2^{k+k'}$, $z_c$ has to be larger than $h_2$ in order for 
the l.h.s of \contint\ to be finite, hence the model to be well-defined). 
At large $k$, this quantity may be evaluated via the method of Hankel
contours. Indeed, the contour may be deformed as shown in Fig.~\intcont\ so
that the dominant contribution to the integral comes from the vicinity
of $z_c$. Using the scaling
\eqn\sett{z=z_c\left(1+{\delta\over k}\right)\ ,}
the variable $\delta$ is now to be integrated back and forth from
$\infty$ to $0$ (below and above the cut of $F(z)$). Using the
expansion
\eqn\expapm{\lambda_+(z)=\lambda_+(z_c)
\left(1-\mu_+ {\delta \over k}+O\left({1\over k^2}\right)\right)}
with
\eqn\muval{\mu_+= - \left.\left(z {d \ \over dz} \log \lambda_+(z)
\right)\right\vert _{z=z_c}\ ,}
we get at large $k$
\eqn\asymptakk{\eqalign{\sum_{k'\geq 0}A_{k,k'}(h_1,h_2) F_{k'}& \sim
\chi\, \Gamma(1-a) {\lambda_+(z_c)^k \over k^a}
{\sin \pi (1-a)\over \pi}
\int_0^{\infty} d\delta \, \delta^{a-1} \,
{\rm e}^{-\mu_{+}\delta}
\cr
&= \chi {\lambda_+(z_c)^k \over (\mu_{+})^a\, k^a}\ .\cr}}
Here we assumed that $\lambda_+(z_c)>\lambda_-(z_c)$, which is valid
as soon as $h_1>0$. The case $h_1=0$ will be discussed later.
From the fixed point condition, this behavior is transferred into
the asymptotics of $g_k$
\eqn\gkasy{g_k\sim  n\, \chi {\lambda_+(z_c)^k \over (\mu_{+})^a\, k^a}\ .}
By a slight modification of our reasoning, one may easily show
 that this asymptotics also holds in the case $a=2$.
Using again the property that $(g_k)^{1/k}\to 1/(4 R_c)$ for $k\to \infty$,
this yields a first consistency relation
\eqn\consistRR{{1\over 4 R_c}=\lambda_+\left({1\over 4 R(1)}\right)\ .}
This relation generalizes \rigidcond\ which is recovered by
noting that $\lambda_+(z)=h_1^2/z$ in the rigid case.

Repeating the proof of
previous Section, we have here again $R(1)\leq R_c$ and $R(1)=R_c$
if and only if the sequence $(g_k)_{k\geq 1}$ is non-generic critical.
Focusing on such a non-generic critical solution, the relation 
\consistRR\ allows to identify $1/(4 R(1))$ as the fixed point $z^*$ of the 
mapping
$z\mapsto \lambda_+(z)$, namely
\eqn\betaval{{1\over 4\, R(1)}=z^*= h_1+2h_2\ .}
A second consistency relation is now obtained by comparing the prefactors
in \gkasy\ and \gkasymp\ and using the
precise value $\chi=2c R(1)/\sin\pi (a-3/2)$. We eventually deduce the
relation
\eqn\consistn{n= 2 (\mu^*)^a \sin \pi(a-3/2), \qquad
\mu^*=-\left.\left(z {d \ \over dz} \log \lambda_+(z)
\right)\right\vert _{z=z^*}=-\lambda_+'(z^*)\ .}
Now we recall that $z\mapsto \lambda_+(z)$ is an involution
in the vicinity of its fixed point $z^*$, which implies
that $\lambda'_{+}(z^*)=-1$ (since $\lambda_+(z)$ obviously
decreases with $z$), a result which may trivially be checked
by a direct calculation. This fixes $\mu^*=1$ and we
recover {\it the same relation \valn\ as in the rigid case}.

To summarize, the fixed point condition \fixepoint\ is
compatible with having the model subcritical, generic critical
or non-generic critical with $a=2\pm b,\ \pi b =\arccos(n/2)$
as long as $h_1>0$.
The case $h_1=0$ leads to a different relation since in this case
$\lambda_+(z)=\lambda_-(z)=h_2 z/(z-h_2)$ hence the contribution of $\lambda_-(z)$
in \contint\ cannot be neglected. Then Eq.~\asymptakk\ must be
replaced by
\eqn\asymptakkzero{\sum_{k'\geq 0}A_{k,k'}(0,h_2) F_{k'}\sim
2 \chi {\lambda_+(z_c)^k \over (\mu_{+})^a k^a}\ .}
At a non-generic point, we get the same relation $z_c=1/(4 R(1))=z^*$
as before (with now $z^*=2h_2$) and the same value $\mu^*=1$,
but comparing the prefactors as was done above now yields the new relation
\eqn\newdisp{n=\sin \pi(a-3/2)\ , \quad h_1=0}
without the factor $2$ in front. We thus expect that our $O(n)$ loop model
for $h_1=0$ and $0<n<1$ should have the same non-generic critical behavior as
the $O(2n)$ loop model for $h_1>0$.

\newsec{Linear integral equation}
In this Section, we show how to rephrase the fixed point conditions
\fixepoint\ and \fixepointrigid\
as linear integral equations, and then analyze their solutions. We
first focus on the rigid case before addressing the general case of
arbitrary $h_1$ and $h_2$.

\subsec{Derivation of the equation in the rigid case}

We start from the general expression \eqforu\ for the function $u(R)$,
defining implicitly the generating function $R(u)$. Differentiating
\eqforu\ with respect to $R$, we readily get
\eqn\eqforuprime{u'(R)=1 - \sum_{k \geq 1} g_k {2k-1 \choose k} k R^{k-1}.}
Independently we observe that, upon performing the change of variable
$u\to {\tilde R}=R(u)$ in \eqforFk, we have
\eqn\eqforFkR{F_k = {2k \choose k} \int_0^{R(1)} {\tilde R}^k
u'({\tilde R}) d{\tilde R}.}
So far we have only rewritten some equations of Section 3.2, related
to bipartite maps. Let us now combine them with the fixed point condition
\fixepointrigid\ transcribing the gasket decomposition for the $O(n)$
rigid loop model: assuming that $n$, $g$ and $h_1$ are such that the
model is well-defined, we plug \eqforFkR\ into
\fixepointrigid, then the result into \eqforuprime, and obtain a {\it
linear integral equation} for $u'(R)$:
\eqn\lineareq{u'(R)=1-6\, g\, R-{n\over 2}
\sum_{k\geq 1}k {2k \choose k}^2 h_1^{2k}\, R^{k-1} \int_0^{R(1)} {\tilde R}^k
u'({\tilde R}) d{\tilde R}}
(note that ${2k-1 \choose k}={2k \choose k}/2$).  By the change of
variables
\eqn\chvar{x={R\over R(1)}, \quad y={{\tilde R}\over  R(1)}, \quad
f(x)=u'(R), \quad \rho = 6 g\, R(1), \quad \tau= 4 h_1 R(1)\ ,}
the linear integral equation is rewritten in the more compact form
\eqn\linearf{\eqalign{f(x)+{n\over 2 \pi}\int_0^1 &
K_\tau(x,y) f(y) dy= 1-\rho x \qquad (0 \leq x \leq 1)
\cr K_\tau(x,y)= \tau^2 & y\,
\psi(\tau^2 x y), \qquad
\psi(t)= \sum_{k\geq 1} {\pi k\over 16^k} {2k \choose k}^2 t^{k-1}\ .\cr}}
We recognize a {\it Fredholm integral equation of the second kind} 
\HLE\ whose unknown is the function $f$ defined on the range $[0,1]$, and
which depends on the non-negative parameters $n,\rho,\tau$. Note the
consistency relation
\eqn\consistf{\int_0^1 f(x)={1\over R(1)}}
which follows from $1=u(R(1))= \int_0^{R(1)} u'({\tilde R})d{\tilde R}
= R(1)\int_0^1 f(x)dx$.

Let us now discuss the conditions on the function $f$ and the
parameters $\rho$ and $\tau$ that arise from our derivation.  First,
the existence of the inverse function $R(u)$ for $u$ between $0$ and
$1$ implies that $u'(R(1)) \geq 0$ and $u'(R)>0$ for $0\leq R <
R(1)$. This immediately translates into the positivity conditions
\eqn\welldeff{f(1) \geq 0, \qquad f(x)>0\ {\rm for}\ 0\leq x < 1}
with $f(1)=0$ iff the model is critical in the sense of Section
3.2. Furthermore, we also obtain that $\rho \leq 1$, since $u'(R) \leq
1 - 6\, g\, R$ by \lineareq. Second, by the discussion of Section 4.1,
we have $\tau \leq 1$ and the case
$\tau = 1$ corresponds to a non-generic critical point. Note that this
is precisely the range on which the equation \linearf\ is
well-defined. Indeed, since $(\pi k /16^k) {2k \choose k}^2 \to 1$ for
$k \to \infty$, the radius of convergence of $\psi(t)$ is 1, with
$\psi(t) \sim 1/(1-t)$ for $t \to 1$, hence $K_\tau(x,y)$ is a smooth
function of $(x,y) \in [0,1]^2$ except for $\tau=1$ where it has a
polar singularity at $x=y=1$. The integral equation still holds for
$x=1$, which implies that $f(y) \to 0$ for $y \to 1$ sufficiently fast.

Conversely, given $\rho$ and $\tau$ both between $0$ and $1$ and such
that the equation \linearf\ admits a solution $f$ satisfying the
positivity conditions \welldeff, we may return to the original
variables as follows. We first compute $R(1)$ via \consistf, then
deduce $u'(R)$, $g$ and $h_1$ via \chvar, and finally obtain $u(R)$ as
the primitive of $u'(R)$ satisfying $u(0)=0$. Following the steps of
the above derivation backwards, we find that $u(R)$ satisfies \eqforu\
with the $g_k$'s given by \fixepointrigid, so that we are indeed
``solving'' the $O(n)$ rigid loop model. The positivity conditions
ensure that the inverse function $R(u)$, thus the model, are
well-defined. In particular, the generating function $F_p^{\rm
loop}(n;g,h_1,0)=F_p$ is directly expressed from $f$ by
\eqn\Fpfx{F_p = {2p \choose p}
{\int_0^1 x^p\, f(x)\, dx \over \left( \int_0^1 f(x)\, dx \right)^{p+1}}\
.}

In conclusion, we have shown that generating functions for the $O(n)$
rigid loop model may be obtained by solving the linear integral equation
\linearf. In practice, the main difficulty is that the change of
parameters $(g,h_1) \to (\rho,\tau)$ is rather intricate. Nevertheless,
if we were able to compute the function $f$ for arbitrary parameters
$n, \rho, \tau$, then we could deduce $F_p^{\rm loop}(n;g,h_1,0)$
in a parametric form. Before further analyzing the solutions of
\linearf, let us extend our formalism to the non-rigid case.

\subsec{Extension to the non-rigid case}

We may repeat the above analysis in the case of arbitrary values of
$h_1$ and $h_2$, replacing the fixed point condition \fixepointrigid\
by the more general one \fixepoint. The main complication is that an
extra sum over a variable $k'$ is involved. The equation \contint\
allows to rewrite this sum as a contour integral.
Furthermore by \eqforFkR\ we have
\eqn\Fz{F(z)= \int_0^{R(1)}
{u'({\tilde R}) d{\tilde R} \over \sqrt{1 - 4\, z\, {\tilde R}}}.}
Substituting these expressions into \fixepoint, then into
\eqforuprime, we may explicitly evaluate the sum over the variable $k$
and derive a linear integral equation for $u'(R)$. Again a more
compact form is obtained after a suitable change of variables: we
let $x$, $y$, $f(x)$, $\rho$ be as in \chvar\ while we now define
\eqn\newtau{\tau=4R(1)(h_1+2h_2).}
After some work, we arrive at the same form for the linear
integral equation
\eqn\linearfbis{f(x)+{n\over 2 \pi}\int_0^1
K_\tau(x,y) f(y) dy= 1-\rho x}
but with a different kernel $K_\tau(x,y)$, now given by
\eqn\newkernel{K_\tau(x,y)= \oint {d\zeta \over
2 {\rm i}\zeta} {1\over (1-\tau y \zeta)^{1/2}}
{1\over 2}\left(
{ \tau \Lambda_+(\zeta)\over(1-\tau x \Lambda_+(\zeta))^{3/2}}
+{ \tau \Lambda_-(\zeta)\over(1-\tau x \Lambda_-(\zeta))^{3/2}}
\right)}
where we may choose as integration contour the circle $|\zeta|=\zeta_0$
for any $\Lambda_+^{-1}(1/(\tau x))<\zeta_0<1/(\tau y)$, and where we introduced
the rescaled functions
\eqn\Aa{\Lambda_{\pm}(\zeta)={\lambda_\pm\left(\left(h_1+2h_2\right)\zeta\right)
\over h_1+2h_2}\ .}
Note that $\Lambda_+$ and $\Lambda_-$ are functions of the ratio $h_1/h_2$,
and therefore $K_\tau(x,y)$ depends implicitly on this ratio. Despite
the apparent complication in the expression for the kernel
$K_{\tau}(x,y)$, much of the discussion of Section 5.1 can be
generalized to the non-rigid case. Using exactly the same arguments,
we find that:
\item{-} the consistency relation \consistf,
\item{-} the positivity conditions \welldeff, where again $f(1)=0$ iff
the model is critical,
\item{-} the inequality $\rho \leq 1$,
\item{-} the expression \Fpfx\ for the $O(n)$ loop model generating
functions $F_p^{\rm loop}(n;g,h_1,h_2)=F_p$,
\par \noindent
all still hold in the non-rigid case. The discussion of $\tau$, in
view of that of Section 4.2 and of the general expression \newtau,
becomes slightly more involved. We nevertheless find that $\tau$
is still between $0$ and $1$, with $\tau=1$ corresponding to a
non-generic critical point. Furthermore, $K_{\tau}(x,y)$ is a smooth
function of $(x,y) \in [0,1]^2$ except in the case $\tau=1$ where,
remarkably, we obtain the same singular behavior as in the rigid
case, provided that $h_1>0$.  More precisely, when $x$ and $y$ tend to
$1$ (keeping $(1-x)/(1-y)$ finite), we have
\eqn\kernelone{K_1(x,y) \sim {1\over 1-xy}\ .}
For the record, let us briefly explain how
this property results from \newkernel\ by a saddle-point approximation.
The contour integral in the latter equation is dominated, for $x,y \to
1$ by the vicinity of $\zeta=1$ and we set
\eqn\scalzeta{\zeta=1+{\rm i}\, \epsilon\, Z \ ,\quad  x=1-\epsilon\, X\ , \quad
y=1-\epsilon\, Y}
with $\epsilon\to 0$. At leading order in $\epsilon$, we have
\eqn\expA{\Lambda_+(\zeta)=1-\epsilon\, \mu^*\, Z\ +\cdots, \quad
{\rm with}\ \mu^*=-\Lambda'_{+}(1)=-\lambda'_{+}(z^*)=1\ }
using again the ``miraculous'' involutivity of $\lambda_+(z)$ around $z^*$.
We finally obtain the estimate
\eqn\newkernellim{K_1(x,y)\sim {1\over \epsilon} \int_{-\infty}^\infty
{dZ\over 4} {1\over (Y-{\rm i}Z)^{1/2}(X+{\rm i}Z)^{3/2}}=
{1\over \epsilon}{1\over X+Y}\sim {1\over 1-xy}\ .}
In the case $h_1=0$, this estimate must be doubled since $\Lambda_-$ has
then an equal, instead of negligible, contribution. We thus have
$K_1(x,y)\sim 2/(1-xy)$ in this case.

In summary, the generating functions $F_p^{\rm loop}(n;g,h_1,h_2)$
for the $O(n)$ loop model on tetravalent maps may be expressed,
via a change of parameters $(g,h_1,h_2) \to (\rho,\tau,h_1/h_2)$, in
terms of the solution of the linear integral equation \linearfbis.

\subsec{Discussion of the solution of \linearfbis\ and of its singular behavior}

\fig{Qualitative phase diagram of the $O(n)$ loop model in the $(\rho,\tau)$ 
plane
for fixed values of $n$ (between $0$ and $2$) and of
$h_1/h_2$.}{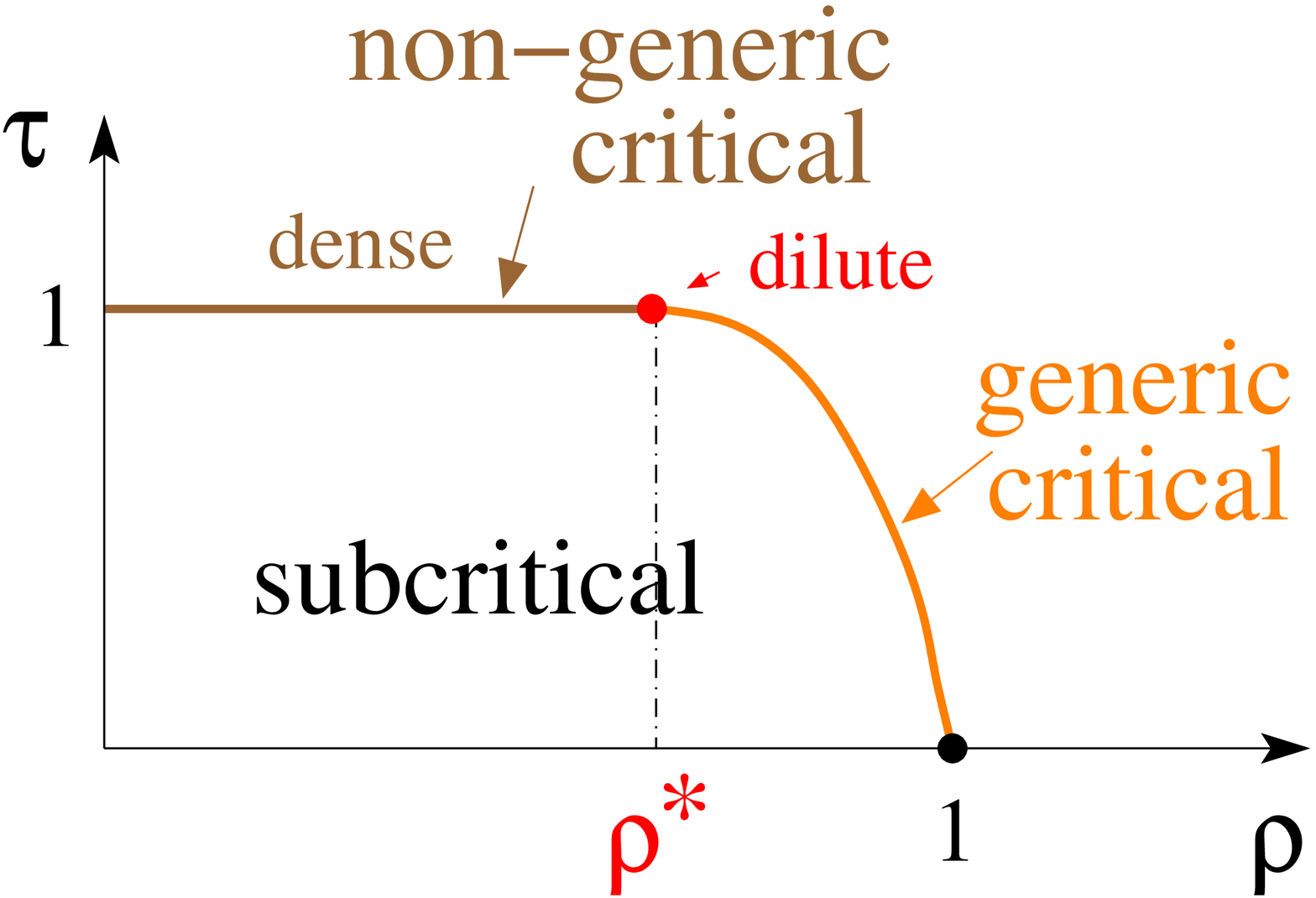}{9.cm}
 \figlabel\phasediagbis

As mentioned previously, Eq.~\linearfbis\ is a Fredholm integral
equation of the second kind for the function $f(x)$. This equation
depends on $n$, $\rho$, $\tau$ and the ratio $h_1/h_2$ as follows:
$-n/(2\pi)$ is the so-called {\it parameter of integral equation},
$\rho$ appears only in the {\it right-hand side} and $\tau$ and
$h_1/h_2$ determine the {\it kernel} \newkernel\ (in particular, for
$h_1/h_2=\infty$, we recover the rigid case of \linearf). Here our
terminology is borrowed from \HLE. In this Section, we shall
assume that $n$ and $h_1/h_2$ are fixed quantities and will look at
the dependence of $f(x)$ on $\tau$ and $\rho$, both varying {\it a
priori} between $0$ and $1$. We shall successively consider the case
$\tau<1$ and the case $\tau=1$, since the latter is special and 
corresponds, as shown before, to non-generic criticality.
Under certain reasonable assumptions, Fredholm theory implies the
qualitative phase diagram of Fig.~\phasediagbis\
in the $(\rho,\tau)$ plane. This picture will be corroborated by the
exact results of Section 6.

\medskip
\noindent {\bf The case $\tau<1$:} here $K_\tau(x,y)$ is a smooth
function of $(x,y)$ varying in the domain $[0,1]^2$, it is thus
square-integrable.  We may then apply Fredholm theory: assuming that
$-(2\pi)/n$ is not a characteristic value of the integral equation (i.e.\ there
exists no non-zero function $f(x)$ such that $f(x)+(n/2\pi)\int_0^1 K_\tau(x,y)
f(y)dy=0$ -- we expect this assumption to be valid when $n$ is between 0 and 2),
Eq.~\linearfbis\ has a unique solution $f(x)$ for each value of
$\rho$. More precisely, by linearity, we have
\eqn\linearho{f(x)=f_1(x)-\rho\, f_{\rm id}(x)}
where $f_1(x)$ (resp.\ $f_{\rm id}(x)$) is the solution of the linear
integral equation obtained by changing its right-hand side into $1$
(resp. $x$). Note that $f_1(x)$ and $f_{\rm id}(x)$ do not depend on
$\rho$ but implicitly depend on $\tau$ as well as $n$ and
$h_1/h_2$. These functions might be expressed, for instance, via
Neumann series at least for $n$ small enough.

However, the solution $f(x)$ does not necessarily satisfy the
positivity conditions \welldeff\ ensuring that the model is
well-defined. In view of \linearho, we conjecture that these
conditions amount to $\rho \leq \rho_c$, where $\rho_c=f_1(1)/f_{\rm
id}(1)$. In particular, when $\rho = \rho_c$, $f(1)=0$ and the model
is critical. We identify
this critical point as generic in the sense of Section 3.2, since we
expect $f'(1)$ hence $u''(R(1))$ to be finite for any $\tau<1$.  As
$\rho_c$ depends on $\tau$, we obtain a generic critical line in the
$(\rho,\tau)$ plane, see Fig.~\phasediagbis.  
Note that, for any values of
$n$ and $h_1/h_2$, the generic critical line starts from
$(\rho=1,\tau=0)$, corresponding to the critical point for pure
quadrangulations without loops.  Indeed, since $K_{0}=0$, we have
$f(x)=1-\rho x$ for $\tau=0$, therefore the critical value of $\rho$
is $1$ (we also recover the known critical values $R(1)=2$ and
$g=1/12$). Moreover, 
for $0<n<2$, we expect the line of generic critical points to
connect continuously with the non-generic critical line $\tau=1$ (to
be discussed below): $\rho_c$ should be positive for all $\tau<1$, and
have a positive limit $\rho^*$ as $\tau \to 1$.

\medskip
\noindent {\bf The case $\tau=1$:} now $K_1(x,y)$ is no longer a
smooth function of $(x,y)$ in the domain $[0,1]^2$, but diverges as
\kernelone\ for $x,y \to 1$. We do not know whether a general theory
applies to such kernels. Nevertheless, from \Tric, we expect that
Eq.\linearfbis\ has, for all $\rho$, a unique continuous solution
satisfying $f(1)=0$. Thanks to this cancellation, the integral
$\int_0^1 K_1(1,y) f(y) dy$ may still be well-defined.

Furthermore, we see that the solution $f(x)$ cannot be regular for $x
\to 1$ (i.e. vanish as an integer power of $x-1$) as otherwise, the
integral in \linearfbis\ would contain singular terms (with
logarithms) which are not present in the r.h.s, regular at $x=1$. This
suggests to assume that the solution behaves as
\eqn\singf{f(x)\sim C (1-x)^\alpha}
for some constant $C$ and some positive non-integral exponent $\alpha$
(the condition $\alpha>0$ ensures both
that $f(1)=0$ and that the integral in \linearfbis\ converges at
$x=1$). Remarkably, $\alpha$ cannot take arbitrary values, but is
related to $n$ via
\eqn\relalpha{n=2 \sin \pi \alpha}
as seen from the following argument. Subtracting to \linearfbis\ its
expression at $x=1$, we get
\eqn\difflinearf{f(x)+{n\over 2 \pi}\int_0^1
(K_1(x,y)-K_1(1,y)) f(y)\, dy= \rho (1- x)}
with $K_1(x,y)-K_1(1,y)\sim - (1-x)y/((1-y)(1-x y))$ when $x$ and $y$
tend to $1$. For $x\to 1$, the dominant singular term in the
integral arises from $y$'s such that $1-y=O(1-x)$. More precisely,
writing $y=1-s(1-x)$, we get at leading order
\eqn\inteval{\left. \int_0^1 (K_1(x,y)-K_1(1,y))
f(y)\, dy\right\vert_{\rm sing.}  \sim - C (1-x)^\alpha
\int_0^\infty {s^{\alpha-1} \over 1+s} ds}
with the right-hand side integral evaluated as $\pi/\sin \pi \alpha$.
As such, this equation holds only for $\alpha<1$ so that the integral 
in the r.h.s is convergent. It remains valid for $\alpha>1$ provided the
integral
is analytically continued, with again the value $\pi/\sin \pi \alpha$
(note that the integral in the l.h.s also contains 
regular terms that, for $\alpha>1$, dominate the leading singular term).
Combining with \singf\ back into \difflinearf, whose r.h.s contains no
singular term, we deduce \relalpha.

Note that, for a given value of $n$ between $0$ and $2$, the possible
$\alpha>0$ satisfying \relalpha\ are of the form $\alpha= 1/2 \pm b +
m$, with $\pi b = \arccos(n/2)$ and $m$ a non-negative integer.  So far
we have only discussed the dominant exponent but, by linearity, all
exponents appearing in the expansion of $f(x)$ at $x=1$ should be
of this form. We therefore expect $f(x)$ to be of the general form
\eqn\genformf{f(x)=(1-x)^{1/2-b}\Phi_{-}(x)+(1-x)^{1/2+b}\Phi_{+}(x)}
where $\Phi_{-}(x)$ and $\Phi_{+}(x)$ are functions with a regular
expansion at $x=1$ (i.e. contain only integral powers of $x-1$). By
considering the expansion of \linearfbis\ at $x=1$ and splitting it
into singular and regular parts (i.e. separating terms with non-integral and
integral exponents), we find that $\Phi_{-}(x)$ and $\Phi_{+}(x)$
should satisfy
\eqn\singulvanish{(1-x)^{1/2\pm b}\Phi_{\pm}(x)+{\cos(\pi b)\over \pi}\left.
\left\{\int_0^1 K_1(x,y) (1-y)^{1/2\pm b} \Phi_{\pm}(y) dy\right\}
\right\vert_{\rm sing.}= 0}
and
\eqn\regimpose{{\cos(\pi b)\over \pi}\left.
\left\{\int_0^1 K_1(x,y) \left((1-y)^{1/2-b} \Phi_{-}(y)
+(1-y)^{1/2+b} \Phi_{+}(y) \right) dy\right\}
\right\vert_{\rm reg.}= 1-\rho x\ .}
Now we expect that $\Phi_{-}(1)$ is not zero generically so that the
leading singularity at $x=1$ is of the form $(1-x)^{1/2-b}$.
Furthermore, $\Phi_{-}(1)$ must be non-negative in order to satisfy
the positivity conditions \welldeff. This situation should hold for
any $\rho<\rho^*$ and corresponds to the dense phase of the
$O(n)$ loop model, see Fig.~\phasediagbis.
At $\rho=\rho^*$, we expect $\Phi_{-}(1)=0$ and
$\Phi_{+}(1)>0$ so that, at this special point, the leading
singularity becomes of the form $(1-x)^{1/2+b}$, corresponding now to
the dilute $O(n)$ loop model. For $\rho>\rho^*$, we expect
$\Phi_{-}(1)<0$ so that the model is ill-defined.
That the critical condition $f(1)=0$
for $\tau<1$ coincides when $\tau \to 1$ with the condition
$\Phi_{-}(1)=0$ is rather natural and makes us believe that the
transition from the line of generic critical points to that of
non-generic ones should be continuous. Note that having an effective
value of $\alpha$ larger than $1$ for the leading singularity simply
corresponds to a situation where both $\Phi_{-}(1)$ and $\Phi_{+}(1)$
would vanish. This may occur only for a particular class of
right-hand sides in the integral equation and
it is not expected in the present case.

Finally, recalling Section 3.3, we see that the behavior \singf\ for
$f(x)=u'(R)$ corresponds precisely to the expansion \expansing\ for
$u(R)$. In particular, we identify
\eqn\alphaa{a=\alpha+{3\over 2}}
and the relation \relalpha\ is nothing but \valn.

\newsec{Exact phase diagram of the rigid loop model}

In this Section, we concentrate on the rigid loop model and explain
how to derive its exact phase diagram.

\subsec{Equations for the resolvent}

The starting point is a linear integral equation, not for the function
$u'(R)$ as in Section 5.1, but for the resolvent $W(\xi)$ defined as
in \resolvdef. Those two quantities are related via
the general formula \resolvu.  Moreover, substituting the fixed point condition
\fixepointrigid\ for the rigid loop model into \eqforu, we have for $R<R(1)$
\eqn\ufromW{\eqalign{u(R)&=R-3 g R^2-{n\over 2}
 \sum_{k \geq 1} {2k \choose k} h_1^{2k} R^{k} F_k \cr
&= R-3 g R^2-{n\over 2}   \oint {d\xi W(\xi) \over 2{\rm i}\pi}
\left({1\over \sqrt{1-4 h_1^2 R\, \xi^2}}-1\right)\ }}
where the contour of integration is, say, the circle of radius
$\gamma=2 \sqrt{R(1)}$ (note that $4 h_1^2 R \gamma^2$ is always
smaller than 1 by Eq.~\rigidcond). Upon differentiating
with respect to $R$, and substituting into \resolvu, we obtain
\eqn\eqforW{\eqalign{W(\xi)
  &= S(\xi) - {n \over \xi}  \oint {d\xi' W(\xi') \over 2 {\rm i} \pi}
  \int_0^{R(1)} dR \left( 1 - {4 R \over \xi^2} \right)^{-1/2} h_1^2 \xi'^2
  \left( 1 - 4 h_1^2 \xi'^2 R \right)^{-3/2} \cr
  &= S(\xi) - {n \over 2 \xi}  \oint {d\xi' W(\xi') \over 2 {\rm i} \pi}
  {h_1^2 \xi'^2 \over h_1^2 \xi'^2 - 1/\xi^2}
   \left( \sqrt{ 1 - 4 R(1)/\xi^2 \over 1 - 4 R(1) h_1^2 \xi'^2} -1 \right) \cr
}}
where
\eqn\eqforS{S(\xi) = {1 \over 2} \left( \xi - g \xi^3 - \xi ( 1 -
2 g R(1) - g \xi^2) \sqrt{1 - 4R(1)/\xi^2} \right)}
corresponds to the first two terms in the r.h.s of \ufromW.

Eq.~\eqforW\ is a linear integral equation for the resolvent which
rephrases that of Section 5.1 for $f(x)=u'(R)$.
It implies a
simpler functional equation for $W(\xi)$ as follows: for $\xi \in
[-\gamma,\gamma]$, we have \eqn\Scut{S(\xi + {\rm i} 0) + S(\xi - {\rm
i} 0) = \xi - g \xi^3} so that, from \eqforW, we may write
\eqn\Wcutinter{W(\xi + {\rm i} 0) + W(\xi - {\rm i} 0) = \xi - g \xi^3
+ {n \over \xi} \oint {d\xi' W(\xi') \over 2 {\rm i} \pi} {h_1^2
\xi'^2 \over h_1^2 \xi'^2 - 1/\xi^2}.}  The latter integral may be
evaluated by the residue theorem: the integrand has poles at $\xi' =
\pm (h_1 \xi)^{-1}$, each with residue $(h_1 \xi)^{-1} W((h_1
\xi)^{-1})/2$, and a pole at $\xi'=\infty$ with residue
$-F_0=-1$. Hence the resolvent satisfies the functional equation
\eqn\Wcut{W(\xi + {\rm i} 0) + W(\xi - {\rm i} 0) = \xi - g \xi^3 + {n
\over \xi} - {n \over h_1 \xi^2} W \left( {1 \over h_1 \xi }
\right),\qquad \xi \in [-\gamma,\gamma].} 
Note the similarity with Eq.\RHprobW\ for the different model of
Section 1.2. Equation \Wcut\ can also be obtained as a consequence of
loop equations in the matrix model formulation of our $O(n)$ model
[\xref\GBThese, Eq.(V-22)].

Solving \eqforW\ boils down to finding a solution of \Wcut\ which is
bounded, odd in $\xi$ and such that $W(\xi) \sim 1/\xi$ as $\xi \to
\infty$. As we shall see, these requirements fix $W(\xi)$ completely.
By linearity, we may write
\eqn\Wlin{W(\xi) = W_{\rm part}(\xi) + W_{\rm hom}(\xi)\ ,}
where $W_{\rm part}(\xi)$ is the easy particular solution of \Wcut
\eqn\Wpart{W_{\rm part}(\xi) = { 2 ( \xi - g \xi^3 ) -
n ( {1 \over h_1^2 \xi^3 } - {g \over h_1^4 \xi^5}) \over 4 - n^2} +
{n \over (2 + n) \xi}}
and where $W_{\rm hom}(\xi)$ is now an odd solution of the homogeneous equation
\eqn\Whom{W_{\rm hom}(\xi + {\rm i} 0) + W_{\rm hom}(\xi - {\rm i} 0) +
{n \over h_1 \xi^2} W_{\rm hom} \left( {1 \over h_1 \xi } \right)=0\ .}
The condition that $W(\xi)\sim 1/\xi$ for $\xi\to\infty$, and that $W$ is
bounded amounts to demanding that
\eqn\Whomexp{\eqalign{W_{\rm hom}(\xi) & = {2 g \over 4 - n^2} \xi^3 -
{2 \over 4 - n^2} \xi + {2 \over 2 + n} \xi^{-1} + O(\xi^{-2}),\quad \xi
\rightarrow \infty \cr
W_{\rm hom}(\xi) & = {-n \over 2}\left({2 g \over 4 - n^2} \xi^{-5} - {2 \over 4
- n^2} \xi^{-3} + {2 \over 2 + n} \xi^{-1} + O(1)\right),\quad \xi \rightarrow
0\ .\cr} }
We shall give in Section 6.3 the general expression for $W_{\rm hom}(\xi)$
when $4h_1R(1) < 1$, which corresponds to a subcritical or generic critical
situation. Something special happens on the non-generic critical line
$4h_1R(1) = 1$ because then, the cut $[-\gamma,\gamma]$ collides with its
image under $\xi \mapsto 1/(h_1\xi)$ (see Fig.10 below). As we shall now see,
$W_{\rm hom}(\xi)$ has a simple expression along this line.

\subsec{Non-generic critical line}

As seen in Section 4.1, the non-generic critical line is characterized by
$\tau = 4 h_1 R(1)=1$ (or equivalently $R(1)=R_c$). This implies
that the extremity of the cut $\gamma=2\sqrt{R(1)}=1/\sqrt{h_1}$ 
is a fixed point of
$\xi \mapsto 1/(h_1 \xi)$. It is then possible to guess the general solution of
the homogeneous equation \Whom\ which is odd in $\xi$, namely:
\eqn\Whomcritng{W_{\rm hom}(\xi)= \left(B(\xi) - {\gamma^2 \over \xi^2}
\,B\left({\gamma^2\over \xi}\right)\right)\left({\xi-\gamma\over \xi+\gamma}
\right)^b -\left(B(-\xi) - {\gamma^2 \over \xi^2}\,B\left(-{\gamma^2\over \xi}
\right)\right)\left({\xi+\gamma\over \xi-\gamma} \right)^b}
where $B(\xi)$ is some arbitrary analytic function and where, again,
$\pi b=\arccos(n/2)$. That this form satisfies \Whom\ can be checked
directly and one can prove that it describes all the solutions. 
This in turn leads to a spectral density supported on $[-\gamma,\gamma]$:
\eqn\rhocritng{\rho(\xi)\!=\!-{\sin(\pi b)\over \pi}\!\left(\!
\left(B(\xi)\!-\!{\gamma^2 \over \xi^2}\,B\left({\gamma^2\over \xi}\right)\!
\right)\left({\gamma\!-\!\xi\over \gamma\!+\!\xi}\right)^b\!+\!\left(B(-\xi)
\!-\!{\gamma^2 \over \xi^2}\,B\left(-{\gamma^2\over \xi}\right)\!\right)
\left({\gamma\!+\!\xi\over \gamma\!-\!\xi}\right)^b\!\right).}
The requirement that $W_{\rm hom}$ is holomorphic in 
${\bf C} \setminus [-\gamma,\gamma]$ imposes that $B$ is an entire function. 
To satisfy \Whomexp, $B(\xi)$ must be a polynomial of degree $3$, 
whose four coefficients are determined from
\eqn\Bdeterm{B(\xi)\left({\xi-\gamma\over \xi+\gamma} \right)^b
= {g \over 4 - n^2} \xi^3 - {1\over 4- n^2} \xi +
{1 \over 2 + n} \xi^{-1}
+ O(\xi^{-3}) \qquad {\rm as}\ \xi \to \infty\ . }
Note that we could a priori imagine terms of order $\xi^2$ and $\xi^0$
in this expansion since they would be canceled by parity in the expansion
of $W_{\rm hom}(\xi)$ at large $\xi$. However, such terms would create poles for $\rho(\xi)$ at $\xi=0$, which are not allowed because $\rho$ should be integrable.
The condition \Bdeterm\ amounts to five equations: four of them fix the
coefficients of $B(\xi)$ and the last one yields some additional relation
between $g$ and $h_1$
which is nothing but the equation for the non-generic critical line.
We find explicitly
\eqn\valB{B(\xi)={g\over 4-n^2} \left(\xi^3+2 b \gamma \xi^2
+2 b^2 \gamma^2 \xi + {2\over 3}(b+2 b^3)\gamma^3 \right) - {1\over
4-n^2} (\xi +2 b \gamma)}
while the equation for the non-generic critical line reads
\eqn\ngeneq{g={3\over 2+b^2}\left(h_1- {2-n\over 2b^2} h_1^2\right)\ .}

In order for the expression \Whomcritng\ to be consistent with
the positivity of $u'(R)$, the spectral
density \rhocritng\ must be positive on $]-\gamma,\gamma[$. In particular,
expanding $\rho(\xi)$ for $\xi\to \pm \gamma$, we have
\eqn\rhoexp{\eqalign{\rho(\xi)={\sin(\pi b)\over \pi} &\left( 4^b
\left(B(-\gamma)-\gamma B'(-\gamma)\right)
\left(1-{\xi^2\over \gamma^2}\right)^{1-b}\right.\cr &
\left.+
 4^{-b}
\left(B(\gamma)+\gamma B'(\gamma)\right)
\left(1-{\xi^2\over \gamma^2}\right)^{1+b}\right) +O\left(1-{\xi^2\over
\gamma^2}\right)^{2-b}\ .\cr}}
Demanding the positivity of $\rho(\xi)$ requires
$B(-\gamma)-\gamma B'(-\gamma)\geq 0$, which yields the condition
\eqn\condcrit{g\leq {3 h_1\over 2(b^2-2 b+3)}\ .}
The non-generic critical line \ngeneq\ cannot extend outside of the region
defined by \condcrit\ and therefore ends when it hits the boundary of
this region, i.e. at a point $(g^*,h_1^*)$ with
\eqn\ghonestar{g^*=
{3 b^2 (2-b)^2 \over 2(2-n)(b^2-2 b+3)^2}
 \ , \quad h_1^*= {b^2 (2-b)^2 \over (2-n)(b^2-2 b+3)}\ .}
As shown in Section 6.4, the critical line becomes generic beyond that point.
When $(g,h_1)=(g^*,h_1^*)$, we may check that the coefficient of the first subleading term is positive, i.e. $B(\gamma)+\gamma B'(\gamma)> 0$.
Comparing \rhoexp\ with \rhosing, we therefore read the value of the exponent $a$,
namely $a=2-b$ for $g<g^*$ and $a=2+b$ at $g=g^*$.

\subsec{General expression of the resolvent}

Eq.~\Wcut{} can be recast more elegantly in terms of the differential form
$\omega(\xi) = W_{\rm hom}(\xi)\,d\xi$, namely:
\eqn\RHdiff{
\forall \xi \in [-\gamma,\gamma],\qquad \omega(\xi + {\rm i}0) + 
\omega(\xi - {\rm i}0) -
n\,\omega(s(\xi)) = 0\ ,}
where $s$ is the involution $\xi \mapsto 1/(h_1\xi)$. Note that
$(s(\xi))^2=1/\lambda_{+}(1/\xi^2)$, where $\lambda_+$ is the involution
of Section 2.3, specialized to the rigid case. We underline the similarity of
Eq.~\RHdiff\ with that relevant in the $O(n)$ model discussed in Section 1.2,
where loops visit only
vertices of degree $3$: in this case, we had a different involution
$\xi\mapsto {\tilde h}^{-1} - \xi$. So, the techniques already developed for
the $O(n)$ model where loops visit only vertices of degree $3$
[\xref\EK,\xref\GBThese] can be applied to Eq.~\RHdiff\ with few modifications.

\fig{The cut of $W(\xi)$, or equivalently of $W_{\rm hom}(\xi)$ (solid line)
and its image (dashed line) under the involution
$\xi\mapsto 1/(h_1\xi)$.}{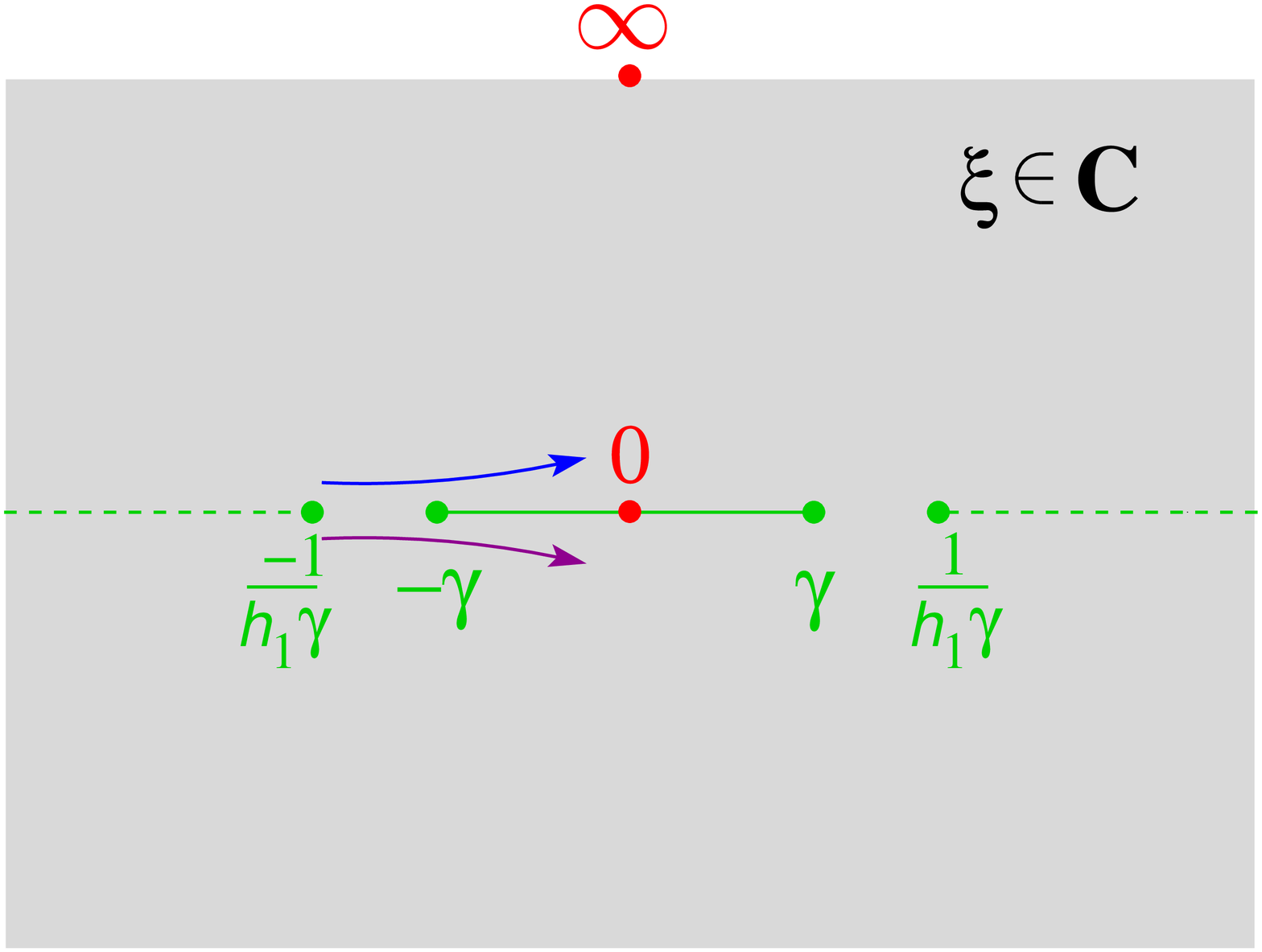}{8.cm}
\figlabel\cuts

\fig{Construction of the elliptic parametrization $v(\xi)$ of Eq.~(6.20),
which depends on the path followed from $-(h_1\gamma)^{-1}$ to $\xi$. The 
upper and lower half-planes map respectively to the left and right rectangles
(whose union is denoted by ${\cal V}$).}{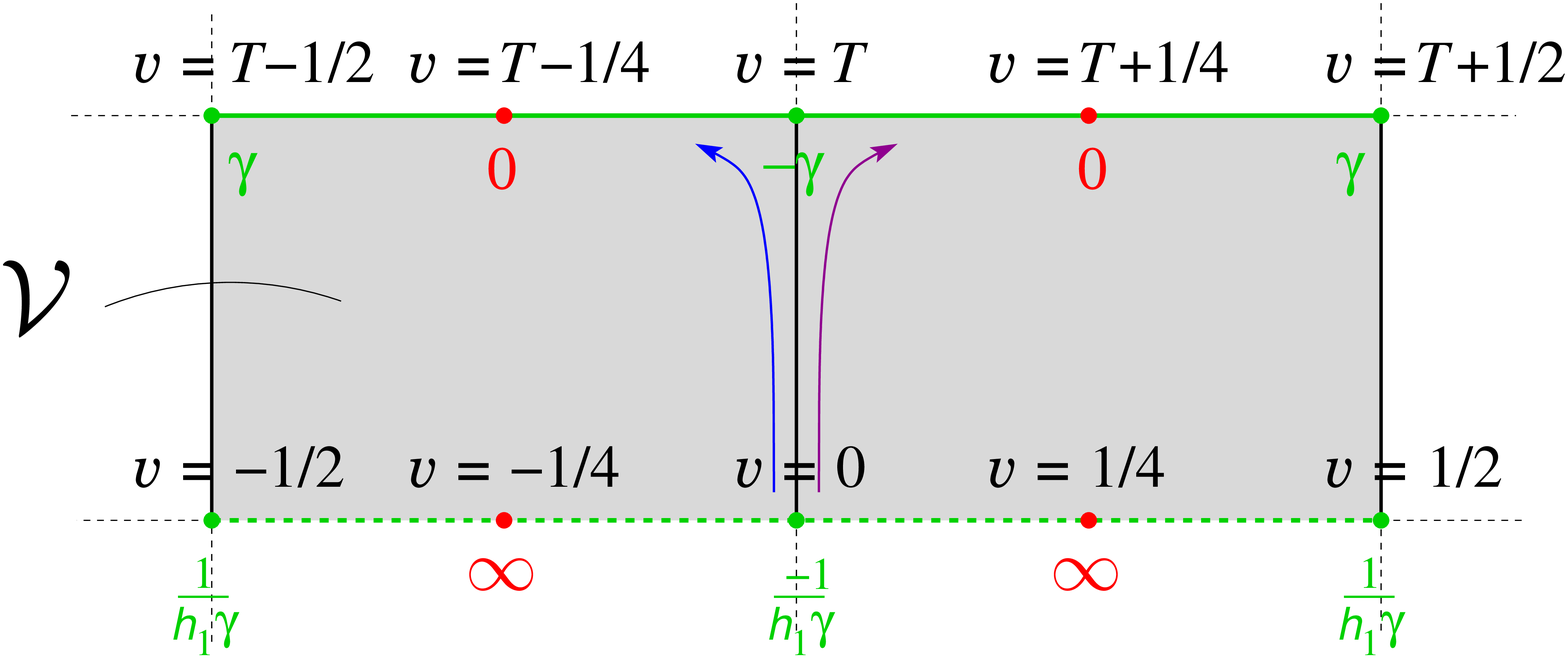}{12.cm}
\figlabel\vplane

The general strategy is to introduce a parametrization which opens the cut
$[-\gamma,\gamma]$ as well as its image
$]-\infty,-(h_1\gamma)^{-1}]\cup[(h_1\gamma)^{-1},+\infty[$ under the
involution (see Fig.~\cuts), for instance:
\eqn\paramdef{
v(\xi) = V\,\int_{-(h_1\gamma)^{-1}}^{\xi} {d\eta \over \sqrt{-(\eta^2 -
\gamma^2)(\eta^2 - (h_1\gamma)^{-2})}}\ .}
The new variable $v$ can be expressed in terms of Jacobi elliptic functions
and is a multivalued function of $\xi$, depending on the path followed in the
complex plane from the origin $(h_1\gamma)^{-1}$ to $\xi$. Conversely, one
may view $\xi$ as a function of $v \in {\bf C}$. We choose the constant $V$ by
demanding that $v((h_1 \gamma)^{-1})=-1/2$ when following a path with small positive imaginary part from $-(h_1\gamma)^{-1}$ to $\infty$ and then back
to $(h_1\gamma)^{-1}$. This leads to
\eqn\cteparam{V = {{\rm i} \over 4h_1\gamma K(h_1\gamma^2)}\ ,}
where $K$ is the complete elliptic integral. We then denote $T = v(-\gamma)$
(by the most straightforward path), and the fact that the square root
discontinuity is included in the real axis implies that $T = {\rm i}\, |T|$.
The function $\xi(v)$ is easily seen to have the following properties:
\eqn\propxi{\eqalign{&\xi(v + 1) = \xi(v)\ ,\quad  \xi(-v) = \xi(v)\ ,\cr
& \xi(v + 1/2) = -\xi(v)\ , \quad \xi(T-v)= \left(h_1\xi(v)\right)^{-1}\ ,\cr }}
from which one can deduce also the properties of its derivative $\xi'(v)$.
For bookkeeping, we mention the expansion when $v \rightarrow v_{\infty}=-1/4$:
\eqn\expxi{\xi(v) = {\Xi_{-1} \over v - v_{\infty}} + \Xi_{1}(v - v_{\infty})
+ O\big((v - v_{\infty})^3\big)}
with
\eqn\expxs{\eqalign{& \Xi_{-1} = {\rm i}V = {-1\over 4h_1\gamma K(h_1\gamma^2)}
= {{\rm i}\,T \over h_1\gamma K'(h_1\gamma^2)} \cr &
\Xi_{1}\Xi_{-1}={1 \over 6}\left(\gamma^2 + (h_1\gamma)^{-2}\right)\ ,\cr }}
where we denote $K'(\tau) = K(\sqrt{1 - \tau^2})$. The image of the points of
${\bf C}\setminus\big([-\gamma,\gamma]\cup]-\infty,-(h_1\gamma)^{-1}]
\cup[(h_1\gamma)^{-1},+\infty[\big)$ reached by a path which does not cross
the segment $[\gamma, (h_1 \gamma)^{-1}]$ (for instance a straight path)
is the domain ${\cal V} = \{v:\ {\rm Re}[v] \in ]-1/2,1/2[\ ,{\rm Im}[v]
\in ]0,|T|[\}$, as shown in Figs.~\cuts\ and \vplane. Let us define
\eqn\varpidef{\varpi(v) = W_{\rm hom}(\xi(v))\xi'(v)}
which is an analytic function on ${\cal V}$. Since $W_{\rm hom}(\xi)$ has no
discontinuity when $\xi \in ]-\infty,-(h_1\gamma)^{-1}]\cup[(h_1 \gamma)^{-1}
,\infty[$, $\varpi(v)$ takes
opposite values when $v \leftrightarrow -v$ along the segment
$[-1/2,1/2]$. Thus, $\varpi$ can be extended to an analytic function defined
on ${\cal V}_2 = {\cal V}\cup (-{\cal V})$ by setting
\eqn\nodisca{\forall v \in -{\cal V},\qquad \varpi(v) = -\varpi(-v)\ .}
Likewise, the absence of discontinuity along $\xi \in [\gamma,(h_1\gamma)^{-1}]$
allows to extend $\varpi$ as an analytic function defined on the strip
${\cal V}_{\rm strip} = \bigcup_{m \in {\bf Z}} \ ({\cal V}_2 + m)$ by setting
\eqn\nodiscb{\forall v \in ({\cal V}_2 + m),\qquad \varpi(v) = \varpi(v - m)\ .}
The top boundary of this strip, ${\cal V}_{\rm cut}=\{v:\ {\rm Im}[v] = |T|\}$,
maps to points $\xi \in [-\gamma,\gamma]$. Eventually, looking at Eq.~\propxi\
and using Eq.~\nodisca, Eq.~\RHdiff{} turns into:
\eqn\omegaeq{\forall v \in {\cal V}_{{\rm cut}},\qquad \varpi(v) +
\varpi(v - 2T) - n\varpi(v - T) = 0\ .}
This allows to extend recursively and without ambiguity $\varpi$ as an analytic
function on ${\bf C} = \bigcup_{m \in {\bf Z}} ({\cal V}_{\rm strip} + 2mT)$.
For instance, for all $v\in {\cal V}_{\rm strip} + 2T$, $\varpi(v)$
can be defined as $n \varpi(v-T)-\varpi(v-2T)$, noticing that $v-T$
and $v-2T$ belong to ${\cal V}_{\rm strip}$.
Since \nodisca, \nodiscb{} and \omegaeq{} are linear analytic 
relations between analytic
functions, they are now valid for any value $v \in {\bf C}$.

To summarize, we reduced the problem to that of finding an analytic function
$v \mapsto \varpi(v)$, which is odd, $1$-periodic, and satisfies:
\eqn\varpieq{\varpi(v - 2T) - n\varpi(v - T) + \varpi(v) = 0\ .}
Eq.~\Whomexp{} demands that $\varpi(v)$ behaves as:
\eqn\varpiexpi{
\varpi(v)\!=\!-{2g\,\Xi_{-1}^4 \over 4\!-\!n^2}\,{1 \over (v\!-\!v_{\infty})^5}
\!+\!{2\,\Xi_{-1}^2 \over 4\!-\!n^2}\left[1\!-\!{g \over 3}\big(\gamma^2\!+
\!(h_1\gamma)^{-2}\big)\right]{1 \over (v\!-\!v_{\infty})^3}\!-\!{2 \over 2\!+\!n}
{1 \over v\!-\!v_{\infty}}\!+\!O(1)}
when $v \rightarrow v_{\infty} = -1/4$, and:
\eqn\varpiexpis{\varpi(v)\!={n\,g\Xi_{-1}^4 \over 4\!-\!n^2
}{1 \over (v\!-\!v_{0})^5}\!-\!{n\,\Xi_{-1}^2 \over 4\!-\!n^2}\left[1\!-\!
{g \over 3}\big(\gamma^2\!+\!(h_1\gamma)^{-2}\big)\right]{1 \over (v\!-\!v_{0})^3}
\!+\!{n \over 2\!+\!n}\,{1 \over v\!-\!v_{0}}\!+\!O(1)}
when $v \rightarrow v_0 = T - 1/4$.
We now assume $n \notin \{0,2\}$, such that $e^{2{\rm i}\pi b} \neq 1$.
Defining 
\eqn\varpipmdefs{\varpi_\pm(v)={\varpi(v-T)-e^{\mp{\rm i}\pi b}\varpi(v)
\over e^{\pm{\rm i}\pi b}-e^{\mp{\rm i}\pi b}}\ , }
we have
$\varpi(v) = \varpi_+(v) + \varpi_{-}(v)$, while, from the $1$-periodicity of $\varpi$
and from \varpieq, $\varpi_{\pm}$ must satisfy:
\eqn\omegatrans{ \varpi_{\pm}(v + 1) = \varpi_{\pm}(v),\qquad \varpi_{\pm}
(v + T) = e^{\pm {\rm i}\pi b}\varpi_{\pm}(v).}
Such functions are generalizations of elliptic functions, and they can be
constructed  by taking appropriate ratios of the Jacobi theta function of nome
$q = e^{{\rm i}\pi T}$. Let us just state the existence of a unique analytic
function $\zeta_{b}$, which satisfies
\eqn\zetaqh{\zeta_b(v + 1) = \zeta_b(v),\qquad \zeta_b(v+T)=e^{{\rm i}\pi b}
\zeta_b(v)}
and has a unique pole when $v = 0\,{\rm mod}\,{\bf Z} \oplus T{\bf Z}$
, which is simple, and is such that $\zeta_b(v) \sim 1/v$ when
$v \rightarrow 0$. Its construction and main properties are listed in
Appendix B. One may generate functions satisfying \omegatrans{} with poles
of higher degree at $v = 0$ by considering the derivatives of $\zeta_b$
or of $\zeta_{-b}(v)=-\zeta_b(-v)$, and put this pole at any given point $w$ by shifting the argument $v$ to
$v - w$. Since holomorphic functions satisfying \omegatrans{} must vanish
identically, one may determine $\varpi$ by matching the divergent behavior at
its poles with a linear combination of the previous functions.
This leads eventually to:
\eqn\ofinal{\varpi(v) = -{1 \over 2 + n}{\cal D}\left\{\zeta_{b}(v - 1/4) +
\zeta_b(v + 1/4) - \zeta_b(- v + 1/4) - \zeta_b(-v - 1/4)\right\}}
where ${\cal D}$ is the differential operator:
\eqn\calD{{\cal D} = {g\,\Xi_{-1}^4 \over 24(2 - n)} \partial_v^4 + {\Xi_{-1}^2
\over 2(2 - n)}\left[-1 + {g \over 3}\big(\gamma^2 + (h_1\gamma)^{-2}\big)
\right]\partial_v^2 + 1\ .}
Then, the spectral density is given by:
\eqn\spdens{\eqalign{\rho(\xi(v))d\xi(v) & = {\xi'(v) \over 2{\rm i}\pi}
\left(W_{{\rm hom}}(\xi(2T - v)) - W_{{\rm hom}}(\xi(v))\right) dv \cr
& = -{dw \over 2\pi}\,\sqrt{{2\!-\!n \over 2\!+\!n}}\,{\cal D}
\left\{\zeta_b\!\left(w\!-\!{1\over 4}\right)\!+\!\zeta_b\!\left(w\!+\!
{1\over 4}\right)\!+\!\zeta_b\!\left(\!-w\!+\!{1\over 4}\right)\!+\!
\zeta_b\!\left(\!-w\!-\!{1\over 4}\right)\right\},}}
where we have set $w = T - v$ so that $w \in [0,1/2]$ corresponds to
$\xi \in [-\gamma,\gamma]$. The value of $\gamma = 2\sqrt{R(1)}$ is determined
a posteriori as a function of $g$ and $h_1$ by requiring from \rhodef\ that
$\rho(\xi) \propto \sqrt{\xi \pm \gamma}$ when $\xi \rightarrow \mp\gamma$,
which is equivalent to demanding that $\rho(\xi(v))\xi'(v) = O(w^2)$ when
$w \rightarrow 0$.

\subsec{Generic critical line}

In the solution above, the generic critical line is the relation between $g$
and $h_1$ obtained by demanding that $\rho(\xi) \propto (\xi \pm \gamma)^{3/2}$
when $\xi \rightarrow \mp \gamma$ (see Eq.~\rhosing).
In other words, in the Taylor expansion of $\rho(\xi(v))\xi'(v)$, as given
by \spdens, when $w \rightarrow 0$, the generic critical line is characterized
by the vanishing of the terms of order $1$ and $w^2$. We may write these two
conditions in a parametric way, with parameter $\tau=h_1 \gamma^2=4 h_1 R(1)$, 
as
\eqn\condcrit{\eqalign{
Z_0 + {\Xi_{-1}^2 \over 2(2 - n)}\left(-1 + {g \over 3h_1}
(\tau + \tau^{-1})\right)\,Z_2 + {\Xi_{-1}^4\,g \over 24(2 - n)}\,Z_4 & = 0\ , \cr
Z_2 + {\Xi_{-1}^2 \over 2(2 - n)}\left(-1 + {g \over 3h_1}
(\tau + \tau^{-1})\right)\,Z_4 + {\Xi_{-1}^4\,g \over 24(2 - n)}\,Z_6 & = 0\ , \cr}}
where
\eqn\Zdef{Z_{2j} = \partial^{(2j)}_{w = 0}\left(\zeta_{b}(w - 1/4) +
\zeta_{b}(w + 1/4)\right)\ .}
Note that $Z_{2j}$ depends implicitly on $\tau$ via $\zeta_b$ which
depends on $T$, itself related to $\tau$ via:
\eqn\kdef{\tau = \left({\vartheta_2(0|4T) \over \vartheta_3(0|4T)}\right)^2 = \left({\vartheta_4(0|{-1 \over 4T}) \over \vartheta_3(0|{-1 \over 4T})}\right)^2\ .}
The solution is:
\eqn\syscrit{\eqalign{h_1 & = {1 \over 2(2 - n)\tau[K'(\tau)]^2}\,{\Delta_8
\over \Delta_6 + 4(\tau^2 + 1)[K'(\tau)]^2\,\Delta_4}\ , \cr
g & = {6 \over 2 - n}\,{\Delta_8\,\Delta_4 \over
(\Delta_6 + 4(\tau^2 + 1)[K'(\tau)]^2\,\Delta_4)^2}\ ,}}
where $K'(\tau)=K(\sqrt{1-\tau^2})$ and the $\Delta$'s are 
functions of $\tau$ defined via:
\eqn\deltadef{\Delta_4 = T^{4}(Z_0Z_4 - Z_2^2),\quad \Delta_6 = -T^{6}(Z_0Z_6 -
Z_2Z_4),\quad \Delta_8 = T^{8}(Z_2Z_6 - Z_4^2)\ .}

Those equations become simpler in the neighborhood of the special points $\tau \rightarrow 0$ (i.e. $h_1 \rightarrow 0$), or $\tau \rightarrow 1$ (i.e. $h_1 \rightarrow h_1^*$, the tip of the non-generic critical line). When $h_1 \rightarrow 0$, it is convenient to use the variable $q = e^{{\rm i}\pi T} \rightarrow 0$ for asymptotics, and we find:
\eqn\gpar{\eqalign{g & = {1 \over 12} - {n \over 18}\,q^{4} + {n \over 36}(7 + n)\,q^{8} + O(q^{12})  \cr
h_1 & = {q^2 \over 2} - \left(2 + {n \over 6}\right)q^{6} + \left(7 
+ 2n + {n^2 \over 18}\right)\,q^{10} + O(q^{14}) \cr}}
As expected, the generic critical line meets the critical point of quadrangulations at $(g = 1/12,h_1 = 0)$, and near this point it behaves as:
\eqn\hpar{h_1 = {3 \over 2}\,\sqrt{2 \over n}\,\sqrt{{1 \over 12} - g},\quad g \rightarrow 1/12}
When $\tau \rightarrow 1$, it is convenient to use the variable 
$q' = e^{-{\rm i}\pi/T} \rightarrow 0$ for asymptotics. 
The computation confirms that the generic critical line ends at the 
tip $(g^*,h_1^*)$ of the non-generic critical line found in 
Eq.~\ghonestar, and we find near this point that:
\eqn\eqncritgen{\eqalign{\left({g - g^* \over g^*}\right) &= 
- {2(1 - 2b) \over 2 + b^2}
\,\left({h_1 - h_1^*\over h_1^*}\right) -{(2 - b)^2 \over 2 + b^2}\,
\left({h_1 - h_1^* \over h_1^*}\right)^2 \cr &+ 
{64 (1 + b)\over (1 - b)(2 + b^2)}
\,\left({(2-b)^2(1-b)^2(3-2b+b^2)\over 4b(1+b)^2(2+b^2)}\right)^{1/b}
\left({h_1 - h_1^* \over h_1^*} \right)^{1/b} \cr & 
+o\left({h_1 - h_1^* \over h_1^*}
\right)^{1/b}\cr}}
The first two terms coincide exactly with Eq.~\ngeneq: when 
passing from the generic to the non-generic critical line, both the slope and 
the curvature remain continuous. For $n=1$ (i.e. $b=1/3$), we find a
leading discontinuity in the third derivative of $g$ with respect to $h_1$,
as expected for the Ising model \BOUKA. 

\subsec{Phase diagram}

\fig{The exact phase diagram of the $O(n)$ rigid loop model
in the $(g,h_1)$ plane. It is shown
here for $b=0.3$ (with $n=2 \cos \pi b$) but it is qualitatively the same 
for any value of $b$ between $0$ and $1/2$. 
The critical line separates the region
where the model is subcritical from the region where it is ill-defined.
The type of criticality changes along the line: generic for $g>g^*$,
non-generic for $g<g^*$ with an exponent $a=2-b$ (dense model),
non-generic at $g=g^*$ with an exponent $a=2+b$ (dilute model). The 
line of non-generic critical points is an arc of parabola which we extended
in dashed line for clarity.}{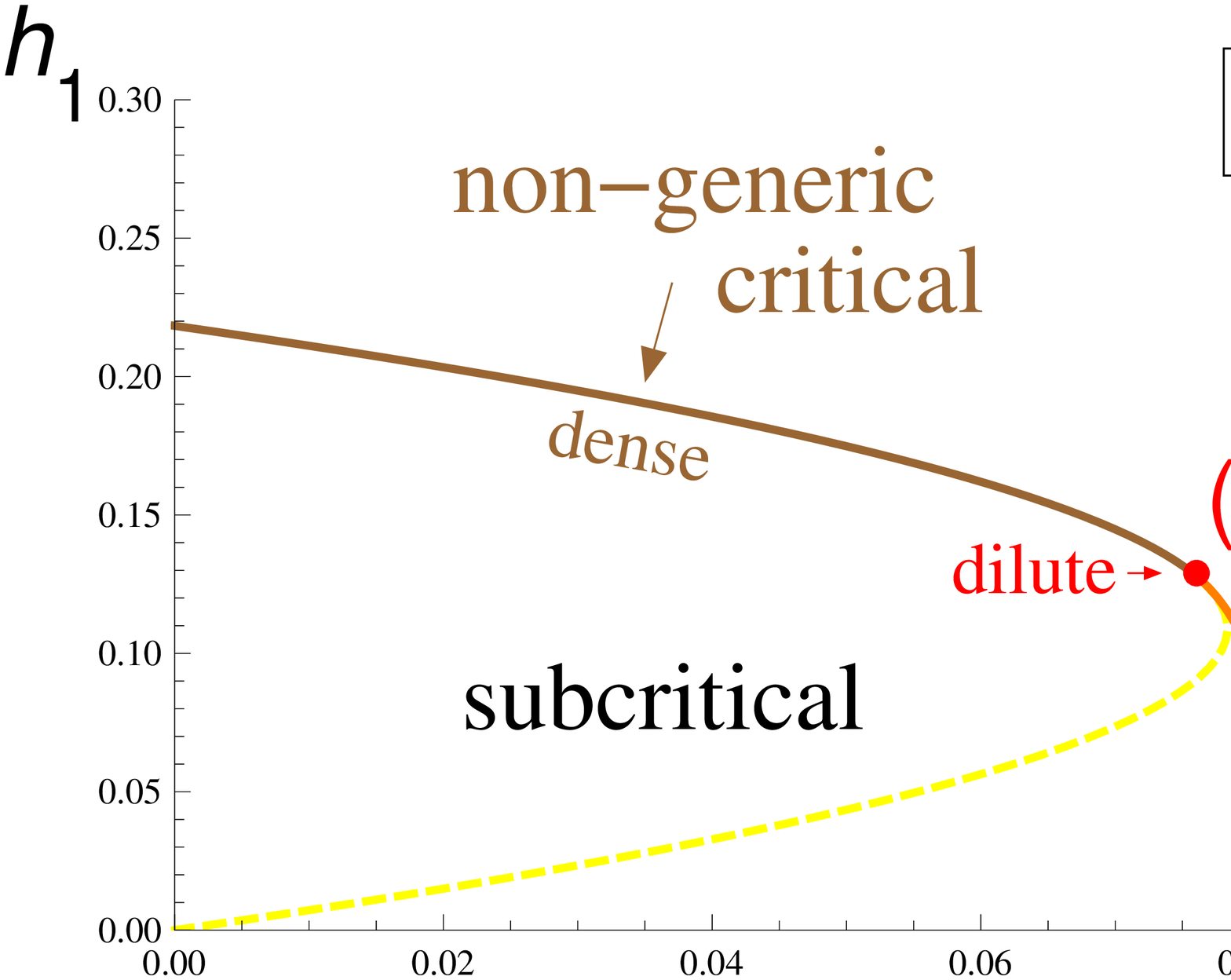}{12.cm}
\figlabel\phasediag
The results above are best summarized in the phase diagram of Fig.~\phasediag.
We have found a line of non-generic critical points given by
the arc of parabola \ngeneq, which links the point 
$(g=0, h_1=2b^2/(2-n))$ to the point $(g^*,h_1^*)$ of Eq.~\ghonestar.
Along this line, the exponent $a$ takes the value $2-b$. At the
terminating point $(g^*,h_1^*)$, $a$ takes instead the value
$2+b$. We then found a line of generic critical points with a more
complicated parametrization \syscrit. 
When $\tau$ decreases from $1$ to $0$,
this line links the point $(g^*,h_1^*)$ to the point 
$(g=1/12,h_1=0)$ describing pure quadrangulations. As just mentioned,
the non-generic and generic critical lines connect with a continuous
slope. Their concatenation forms the line $h_1=h_c(n;g)$ of Section 
4.1 as the model cannot be well-defined above this line.

\fig{The exact phase diagram of the rigid $O(n)$ model in the limit 
$n\to 0$.}{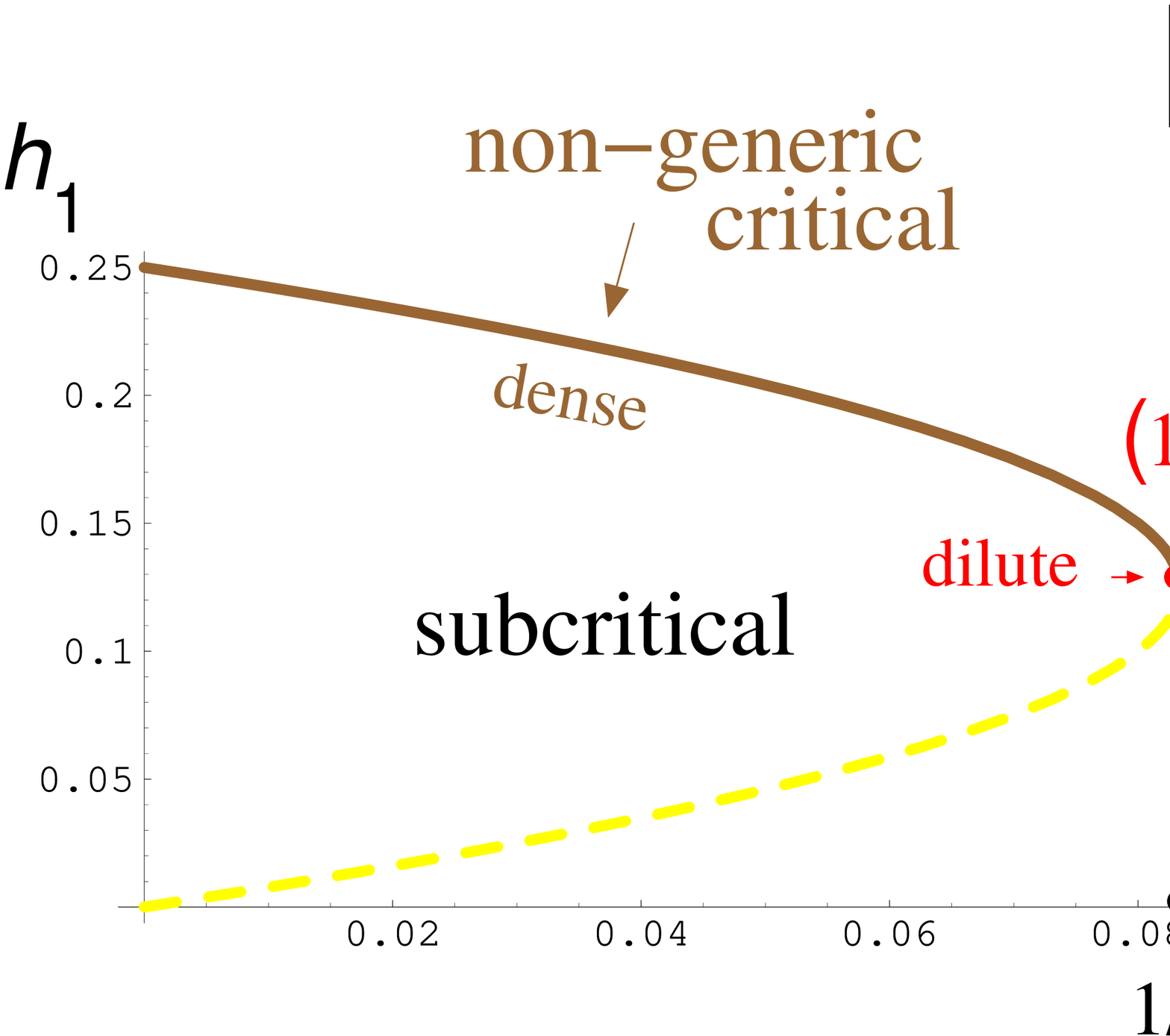}{12.cm}
\figlabel\phasediagzero
Let us finally note that, in the limit $n\to 0$, then $g^* \to 1/12$
and $h_1^* \to 1/8$. The non-generic critical line tends to the
arc of parabola
\eqn\nzeronongen{g={4\over 3} \left(h_1-4 h_1^2\right), \qquad {1\over 8}
\leq h_1 \leq {1\over 4}\ , }
while the generic critical line becomes the vertical segment parametrized
by $g=1/12$, $0\leq h_1 \leq 1/8$ (see Fig.~\phasediagzero). 
This may be understood as follows: 
the small $n$ expansion of 
$F_p^{\rm loop}$ describes quadrangulations equipped with a fixed finite number 
of rigid loops. Using exact enumeration results for quadrangulations
with multiple boundaries, it can be seen that the contribution to 
$F_p^{\rm loop}$ from
loops of large length $2k$ behaves as $(4 h_1 R_Q)^{2k}$, where
$R_Q$ is the generating function $R(1)$ for pure quadrangulations,  
solution of
\eqn\ronequad{R_Q=1+3 g\, R_Q^2\ .}
Having a finite contribution from large loops requires that 
$4 h_1 R_Q \leq 1$. Note that $R_Q$ ranges from $1$ to $2$ 
when $g$ ranges from $0$ to $1/12$. Criticality may be obtained in two 
manners: either we set $4 h_1 R_Q=1$, so that the contribution 
from large loops decays sub-exponentially. Note then the equivalence between
Eqs.~\nzeronongen\ and \ronequad. Or we may set $g=1/12$, so that the pure 
quadrangulations are themselves critical. This last situation requires
$h_1\leq 1/8$, since $R_Q=2$ in this case.

\newsec{Variants of the $O(n)$ loop model}
In this Section, we briefly discuss other versions of the $O(n)$ loop model
on quadrangulations, including models with non-symmetric local weights and
models with restricted loop lengths. More precisely, we concentrate
on non-generic critical points and discuss how the
relation \valn\ is modified in these cases. We finally extend our
results to maps whose faces have arbitrary (but bounded) even degrees.

\subsec{Non-symmetric models}
At this stage, it should be clear to the reader that the relation
\valn\ between the loop weight $n$ and the exponent $a$ only depends
on a few properties of the ring transfer matrix $M(z)$. Denoting by
$\lambda_+(z)$ the largest eigenvalue of $M^2(z)$, we used the estimate
$\sum_{k'\geq 0}A_{k,k'}z^{-k'}\sim \lambda_+(z)^k$ to obtain eventually
the relation
\eqn\valngen{n= 2 (\mu^*)^a \sin \pi(a-3/2), \qquad
\mu^*=-\lambda_+'(z^*)\ ,}
where $z^*$ is the fixed point of the mapping $z\mapsto \lambda_+(z)$.
In the symmetric case discussed so far, this mapping is an involution
in the vicinity of $z^*$, so that $\lambda'_{+}(z^*)=-1$.
Let us now consider a slightly modified, non-symmetric version of our model
defined as follows: the squares of type (c) in Fig.\squares\ come
in two species, those whose two edges not crossed by the loop belong
to the inner contour, and those where these two edges belong to
the outer contour. We may as well view these two species as corresponding
to {\it outward}, resp. {\it inward turns} of the loop at hand.
For instance, the loop in Fig.~\contours\
makes $8$ inward and $6$ outward turns.
Assigning now different weights, say $h_{2,{\rm out}}$ and $h_{2,{\rm in}}$ respectively
to these squares, the transfer matrix $M(z)$ is replaced by
\eqn\newmatr{M(z)\!=\pmatrix{{\displaystyle{h_1 z^{-1/2} \over 1-h_{2,{\rm out}}
z^{-1}}}&
\displaystyle{{1 \over 1-h_{2,{\rm out}} z^{-1}}}
\cr & \cr h_{2,{\rm in}} & 0 \cr}\ .}
The relation \detakk\ becomes
\eqn\detakknonsym{\left( h_{2,{\rm out}}\lambda+ h_{2,{\rm in}} z
- \lambda z \right)^2 - h_1^2 \lambda z =0}
whose largest solution $\lambda_+(z)$ now has a fixed point
at $z^*=h_1+h_{2,{\rm out}}+h_{2,{\rm in}}$.
Since $\lambda$ and $z$ do not play symmetric roles in \detakknonsym,
the mapping $z \mapsto \lambda_+(z)$ is no longer an involution
and we now have a non-trivial value
$\mu^*=(h_1+2h_{2,{\rm out}})/(h_1+2h_{2,{\rm in}})$, leading to the new relation
\eqn\valnnonsym{n=2 \sin \pi (a-3/2)\left(
{ h_1+2h_{2,{\rm out}}\over h_1+2 h_{2,{\rm in}} }\right)^a\ .}
Note in particular that the range of $n$ allowing for non-generic criticality
is modified.
It is instructive to compare this result to that obtained
on a regular square lattice. In this context the parametrization
$n=2 \sin \pi (a-3/2)=2\cos \pi b$ naturally appears in the Coulomb
gas approach to the model \Nien. Many critical exponents of the
model have simple (typically polynomial) expressions in terms of $a$ (or $b$).
On such a regular lattice, we also have
a well-defined notion of exterior and interior of a loop
and we may give different weights to outward and inward turns.
On a regular lattice, there are however $4$ more inward than
outward turns, so that the symmetry of the model may be
restored at the price of a rescaling $n\to n (h_{2,{\rm in}}/h_{2,{\rm out}})^2$
of the weight per loop. This in turn changes the relation
\valn\ into $n=2\sin\pi (a-3/2)(h_{2,{\rm out}}/h_{2,{\rm in}})^2$.

Returning to the $O(n)$ model on quadrangulations, the case $h_1=0$ is,
as before, special since the largest eigenvalue is
degenerate in this case, hence \valnnonsym\ is replaced by
\eqn\valnnonsymbis{n= \sin \pi (a-3/2)\left(
{ h_{2,{\rm out}}\over h_{2,{\rm in}} }\right)^a\ .}

\subsec{Loops with restricted lengths}
Returning to the symmetric case, we may impose some
restriction on the lengths of the loop by demanding
for instance that they be multiples of a fixed integer $N$.
Such a restriction may occur for instance when the loop model
is inherited from some underlying edge coloring problem.
Now the length of a loop whose outer and inner contours
have length $2k$ and $2k'$ is simply $(k+k')$.
The consistency relation \fixepoint\ has to be modified to account
for the new constraint, and we are naturally led to consider now the
quantity
\eqn\akkxN{\sum_{k'\geq 0\atop k+k'=0 \ {\rm mod}\ N}\hskip -20pt
A_{k,k'}(h_1,h_2) z^{-k'}={1\over N}
\sum_{j=0}^{N-1}\left((\omega^j \lambda_+(\omega^{-j} z))^k+(\omega^j
\lambda_-(\omega^{-j} z))^k\right), \qquad  \omega={\rm e}^{2{\rm i}\pi/N}\ ,}
to be estimated as before in the vicinity of $z^*$ and for large $k$.
The above sum behaves as $(1/N)\ (\lambda_+(z^*))^k$ with now a $1/N$ prefactor
provided $\vert \lambda_+(z^*\omega^{-j})\vert < \lambda_+(z^*)$
when $j=1,\cdots, N-1$, which holds for $h_2>0$ (again we also
suppose that $h_1> 0$ to avoid that $\lambda_-(z^*)= \lambda_+(z^*)$).
The correcting factor $(1/N)$ trivially results into a change of the
relation \valn\ into
\eqn\valnmult{n= 2 N \sin \pi(a-3/2)}
(for $h_1,h_2>0$). In particular, imposing an even size for the loops
takes the $O(n)$ loop model in the universality class of the $O(n/2)$
loop model without the parity constraint. This fact was already recognized
in Ref.~\EKbis\ in the slightly different context of loops living on triangles.

In the rigid case $h_2=0$, loops are automatically of even length by
construction. We have
$\omega^j \lambda_+(\omega^{-j} z^*)=\omega^{2j} \lambda_+(z^*)$
and the above sum behaves as $\lambda_+(z^*)^k$, as before, provided $k$ is
a multiple of $N$ for $N$ odd (respectively a multiple of $N/2$ for $N$ even)
while it vanishes otherwise.
For $h_2=0$, the relation \valn\ is therefore unchanged.
Finally, for $h_1=0$, we get $n=N \sin\pi (a-3/2)$ instead.

\subsec{Faces with arbitrary even degrees}
Our results are easily extended to the case of maps whose faces
have arbitrary even degrees, provided these degrees remain bounded,
say by $2M+2$ ($M\geq 1$).
Faces of degree $2m$ not visited by a loop receive a non-negative weight
$g^{(m)}$ ($1\leq m\leq M+1$) and those visited by a loop 
receive a weight $h^{(m_1,m_2)}$ ($m_1,m_2 \geq 0$) if the face
has $m_1$ (resp. $m_2$) incident edges belonging to the outer
(resp. inner) contour of the loop at hand. Since the total degree
of such a face is $m_1+m_2+2$, we will implicitly assume in the
following that $h^{(m_1,m_2)}$ is non zero only if $m_1+m_2\leq 2M$ and
$m_1$ and $m_2$ have the same parity. With these new weights,
the $O(n)$ loop model is now described by the fixed
point condition
\eqn\fixepointgene{g_k=\sum_{m=1}^{M+1}g^{(m)}\delta_{k,m}
+n \sum_{k'\geq 0} A_{k,k'} F_{k'}}
where $A_{k,k'}$ is the generating function for (rooted) rings
(now made of faces with arbitrary even degrees) with sides of lengths
$2k$ and $2k'$. As before, this generating function is best encoded in
the quantity
\eqn\AtoM{\sum_{k'\geq 0} A_{k,k'}z^{-k'} = \tr M(z)^{2k}}
involving a new transfer matrix $M(z)$ of size $2M\times 2M$ now given
by:
\eqn\matgen{M(z)=\pmatrix{
{\sum h^{(1,m_2)}z^{-{m_2\over 2}}\over 1-\sum h^{(0,m_2)}z^{-{m_2\over 2}}}
& {1\over 1-\sum h^{(0,m_2)}z^{-{m_2\over 2}}} & 0 & 0 & \cdots & 0 \cr
\sum h^{(2,m_2)}z^{-{m_2\over 2}} & 0 & 1 & 0 & \cdots & 0  \cr
\sum h^{(3,m_2)}z^{-{m_2\over 2}} & 0 & 0 & 1 & \cdots & 0  \cr
\sum h^{(4,m_2)}z^{-{m_2\over 2}} & 0 & 0 & 0 & \ddots & 0 \cr
\vdots & \vdots & \vdots & \vdots & \ddots & \vdots \cr
\sum h^{(2M,m_2)}z^{-{m_2\over 2}} & 0 & 0 & 0 & 0 & 0 \cr
}\ . }
Here the sum in $\sum h^{(i,m_2)}z^{-{m_2\over 2}}$ runs over $m_2\geq 0$
(and in practice over values of $m_2$ ranging from $0$ to $2M-i$ and
having the same parity as $i$).
As shown in Appendix A, the
eigenvalues $\lambda$ of $M^2(z)$ are the solutions of the 
characteristic equation
\eqn\characeq{\Big((\lambda\, z)^M
-\hskip -5.pt \displaystyle{\sum_{m_1,m_2\geq 0\atop m_1,m_2\ {\rm even}}}
\hskip -10.pt h^{(m_1,m_2)} \lambda^{{2M-m_1\over 2}} z^{{2M-m_2\over 2}}\Big)^2=
\Big(\displaystyle{\sum_{m_1,m_2\geq 0\atop m_1,m_2\ {\rm odd}}}
\hskip -10.pt h^{(m_1,m_2)} \lambda^{{2M-m_1\over 2}} z^{{2M-m_2\over 2}}
\Big)^2\ .}
Repeating the analysis of Section 4, we again find at a non
generic critical point the consistency relation between $n$ and
the exponent $a$ characterizing the large $k$ asymptotics of $F_k$:
\eqn\consistngen{n= 2 (\mu^*)^a \sin \pi(a-3/2), \qquad
\mu^*=-\lambda_+'(z^*)}
in terms of the largest eigenvalue $\lambda_+(z)$ of $M^2(z)$ and its
fixed point $z^*$. Note that this relation holds when the largest
eigenvalue is not degenerate. In a symmetric model, we must set
$h^{(m_1,m_2)}=h^{(m_2,m_1)}$ so that $\lambda$ and $z$ play symmetric
roles in \characeq. This implies as before that the mapping
$z\to \lambda_+(z)$ is an involution in the vicinity of $z^*$ and
that $\mu^*=1$ so that the
simple relation \valn\ is recovered.

This generic relation \consistngen\ is modified whenever $h^{(m_1,m_2)}=0$
for all odd values of $m_1$ and $m_2$. In this case, the r.h.s of
Eq.~\characeq\ vanishes and the largest
eigenvalue is degenerate, resulting in the suppression of the factor
$2$ in \consistngen, namely $n=(\mu^*)^a \sin \pi(a-3/2)$
with again $\mu^*=1$ in the symmetric case.

\newsec{Conclusion}

In this paper, we have shown how to relate a number of $O(n)$ loop models 
to models of bipartite maps. More precisely, we have shown that the gasket
of an $O(n)$ loop model configuration is distributed according to a
Boltzmann ensemble of bipartite maps with appropriate degree dependent
face weights $(g_k)_{k\geq 1}$. Those weights are determined by a fixed
point condition inherited from a bijective decomposition of the
$O(n)$ configurations along the contours of their loops. In particular, the
non-generic (dense and dilute) critical points of the $O(n)$ loop models 
correspond to ensembles of bipartite maps with large faces belonging
to the class considered in \LGM\ with a distribution characterized by some 
exponent $a$ between $3/2$ and $5/2$, related to $n$ generically via
\eqn\valnagain{n=2 \sin \pi (a-3/2)\ .}  
Technically, this formula is one of a number of simple consistency relations
dictated by the fixed point condition at a non-generic critical point. 
Their derivation involves only a 
few properties of a simple transfer matrix $M(z)$ describing the sequence
of faces visited by a loop (the ring). For instance, the constant $z^*$ 
characterizing the exponential decay of the face weights ($g_k\sim (z^*)^k$)
or the exponential growth of $F_k^{\rm loop}$ ($F_k^{\rm loop} \sim (1/z^*)^k$)
is simply obtained as the solution of the equation $z^*=\lambda_+(z^*)$,
where $\lambda_+(z)$ denotes the largest eigenvalue of $M^2(z)$.

Noticeably, the same scheme appears to work also for $O(n)$ loop models
where the loops visit only trivalent vertices or, equivalently on the dual, 
where the ring is made of a sequence of triangles. The ring transfer matrix
reduces in this case to a scalar ${\tilde M}(z)=({\tilde \lambda}_+(z))$, with
\eqn\Mztri{{\tilde \lambda}_+(z)={\tilde h}/(1-{\tilde h}/z)}
(note that we do not square ${\tilde M}$ as contours are not required to 
have even lengths). Here again, the exponential growth factor 
$(1/z^*)=(2{\tilde h})^{-1}$ for ${\tilde F}_p^{\rm loop}$ 
in \asymresolv\ is the solution of  
$z^*={\tilde \lambda}_+(z^*)$, while equation \valnagain\ still holds.

A corollary of our reformulation is that the metric properties of 
the gasket at a non-generic point may be obtained from those of ensembles 
of bipartite maps with large faces. In \LGM, it was shown that these 
maps have a fractal dimension $2a-1$ and one may hope to be able 
to extract the average gasket profile from known expressions for discrete 
distance dependent two-point functions in bipartite maps.

\bigskip
\noindent{\bf Acknowledgments:} The work of G.B. is partly supported 
by the ANR project GranMa ``Grandes Matrices Al\'eatoires'' ANR-08-BLAN-0311-01.

\appendix{A}{Eigenvalues of the transfer matrix for arbitrary even degrees}

Here we consider the $O(n)$ loop model on maps with arbitrary even face degrees,
as defined in Section 7.3. Let us introduce the quantity
\eqn\Sedf{S=\sum_{k\geq 1}{1\over 2k} w^{-k}
\sum_{k'\geq 0} A_{k,k'} z^{-k'}}
where $A_{k,k'}$ enumerates configurations of rooted rings with outer
and inner sides of lengths $2k$ and $2k'$, with the face weights
of Section 7.3. Due to the factor $1/(2k)$, $S$ enumerates {\it unrooted}
ring configurations of arbitrary side lengths, with a weight
$w^{-1/2}$ (resp. $z^{-1/2}$) per edge of the outer (resp. inner) contour.
By a direct calculation, we have
\eqn\Slog{\eqalign{S&=\sum_{k\geq 1}{1\over 2k}\tr(w^{-1} M^2(z))^k\cr
&= -{1\over 2}\tr\log(1-w^{-1}M^2(z)) \cr
&= -{1\over 2} \log \det (1-w^{-1}M^2(z))\ ,\cr}}
where $M(z)$ is the transfer matrix defined in \matgen.
On the other hand, if we denote by $\ell$ the length of a ring, i.e.
its number of faces or equivalently the length of the underlying loop,
the configurations counted by $S$ are simply cyclic sequences of length $\ell$
made of the various squares at hand. We may therefore write
\eqn\Sbis{\eqalign{S&=\sum_{\ell\geq 1}{1\over \ell}
{1\over 2} \Big\{
\Big(\hskip -5.pt \sum_{m_1,m_2\geq 0}
\hskip -10.pt h^{(m_1,m_2)} w^{-{m_1\over 2}} z^{-{m_2\over 2}}\Big)^\ell+
\Big(\hskip -5.pt \sum_{m_1,m_2\geq 0}
\hskip -10.pt h^{(m_1,m_2)} (-1)^{m_1}w^{-{m_1\over 2}} z^{-{m_2\over 2}}
\Big)^\ell
\Big\} \cr &\qquad
-\sum_{\ell\geq 1}{1\over \ell}\Big(
\sum_{m_2\geq 0} \hskip -5.pt h^{(0,m_2)} z^{-{m_2\over 2}} \Big)^\ell
\cr}\ .}
Here the first two terms differ only by the factor $(-1)^{m_1}$,
so that their half-sum selects configuration where the total length of the
 outer contour
is even (and so is that of the inner contour since $m_1$ and $m_2$ have the
same parity whenever $h^{(m_1,m_2)}\neq 0$).
The third term is the $w^{-1}\to 0$ limit
of the first two and is subtracted to account for the fact that there is no
$k=0$ term in $S$.
Summing over $\ell$, this leads to
\eqn\Sfinal{S=-{1\over 2} \log \left( {
\Big(1-\hskip -5.pt \displaystyle{\sum_{m_1,m_2\geq 0}}
\hskip -10.pt h^{(m_1,m_2)} w^{-{m_1\over 2}} z^{-{m_2\over 2}}\Big)
\Big(1-\hskip -5.pt \displaystyle{\sum_{m_1,m_2\geq 0}}
\hskip -10.pt h^{(m_1,m_2)} (-1)^{m_1}w^{-{m_1\over 2}} z^{-{m_2\over 2}}
\Big)
\over \Big(1-
\displaystyle{\sum_{m_2\geq 0}} \hskip -3.pt h^{(0,m_2)} z^{-{m_2\over 2}} \Big)^2  }
\right)}
from which we readily extract the value of $\det(1-w^{-1}M^2(z))$
by comparison with \Slog.
We end up with the final expression
\eqn\dettwo{\det(\lambda-M^2(z))={P(\lambda,z)\over Q(z)} }
where $P(\lambda,z)$ is the polynomial:
\eqn\Pexpr{\eqalign{P(\lambda,z)& = \Big((\lambda\, z)^M
-\hskip -5.pt \displaystyle{\sum_{m_1,m_2\geq 0}}
\hskip -10.pt h^{(m_1,m_2)} \lambda^{{2M-m_1\over 2}} z^{{2M-m_2\over 2}}\Big)\cr
& \qquad \times
\Big((\lambda\,z)^M-\hskip -5.pt \displaystyle{\sum_{m_1,m_2\geq 0}}
\hskip -10.pt h^{(m_1,m_2)} (-1)^{m_1}\lambda^{{2M-m_1\over 2}} z^{{2M-m_2\over 2}}
\Big)\cr}
}
and $Q(z)$ the polynomial:
\eqn\Qexpr{Q(z)=
\Big(z^M-
\displaystyle{\sum_{m_2\geq 0}} \hskip -2.pt h^{(0,m_2)} z^{{2M-m_2\over 2}} \Big)^2
\ .}
Writing $P(\lambda,z)=0$ leads precisely to Eq.~\characeq.

\appendix{B}{Properties of the function $\zeta_b$}

The (first) Jacobi theta function $\vartheta_1(v|T)$ is defined by \WangGuo{}:
\eqn\jactheta{\vartheta_1(v|T) = {\rm i}\sum_{m \in {\bf Z}} (-1)^m\,e^{{\rm i}\pi
(m - 1/2)^2 T + {\rm i}\pi(2m - 1)v}}
The series is absolutely convergent when ${\rm Im}[T]> 0$, and the following
properties can be easily proved:
\eqn\thetaprop{\vartheta_1(v + 1|T) = -\vartheta_1(v|T),\quad
\vartheta_1(v + T|T) = -e^{{\rm i}\pi T - 2{\rm i}\pi v}\,\vartheta_1(v|T)\ .}
The point $v = 0$ turns out to be the unique zero (modulo ${\bf Z} \oplus T
{\bf Z}$) of $\vartheta_1(v|T)$. This function has the following modular
property:
\eqn\thetaprop{\vartheta_1(v|T) = {{\rm i} \over \sqrt{-{\rm i}T}}\,e^{-{{\rm i}\pi
v^2 \over T}}\,\vartheta_1\left({v \over T}\left|\right.{-1 \over T}\right)}
which is useful to relate the limit $|T| \rightarrow 0$ to the limit
$|T'|\rightarrow \infty$, where $T'=-1/T$.
Notice that when
$|T'| \rightarrow +\infty$, we have $q' = e^{{\rm i}\pi T'} \rightarrow 0$,
and therefore the series \jactheta{} gives in a straightforward way the
asymptotics of $\vartheta_1(v|T')$. Other Jacobi theta functions that appear
in the text are:
\eqn\thetao{\eqalign{\vartheta_2(v|T) & = \sum_{m \in {\bf Z}} e^{{\rm i}\pi (m - 1/2)^2 T + {\rm i}\pi(2m - 1)v} \cr
\vartheta_3(v|T) & = \sum_{m \in {\bf Z}} e^{{\rm i}\pi  m^2 T + 2{\rm i}\pi v}\cr
\vartheta_4(v|T) & =  \sum_{m \in {\bf Z}} (-1)^m\,e^{{\rm i}\pi m^2 T  + 2{\rm i}\pi m v}}}

We also introduce the Weierstra{\ss}{} elliptic function:
\eqn\wpdef{\wp(v|T) = {1 \over v^2} + \sum_{(l,m) \in {\bf Z}^2\setminus
\{(0,0)\}} \left({1 \over (v + l + mT)^2} - {1 \over (l + mT)^2}\right)\ .}
The function $\wp$ is even, $1$ and $T$ periodic, and has the following
properties:
\eqn\wpprop{\wp(v|T) \mathop{=}_{v \rightarrow 0} {1 \over v^2} + O(v^2),
\qquad \wp(v|T) = T^{-2}\,\wp\left({v \over T}\left|\right.{-1 \over T}\right)}
Ratios of $\vartheta_1$ may be used to construct functions having prescribed
poles and zeroes, and taking a constant or linear phase when
$v \rightarrow v + T$. For instance:
\eqn\zetadef{\zeta_b(v) = {\vartheta_1(v - b/2|T) \over
\vartheta_1(v|T)}\,{\vartheta_1'(0|T) \over \vartheta_1(-b/2|T)}}
is the unique function which is $1$-periodic, takes a phase $e^{{\rm i}\pi b}$
when $v \rightarrow v + T$, has only a simple pole at
$v = 0\,{\rm mod}\, {\bf Z}\oplus T{\bf Z}$, and is such that
$\zeta_b(v) \sim 1/v$ when $v \rightarrow 0$. The value of the phase under
translation by $T$ implies that $\zeta_b$ has a unique zero
(modulo ${\bf Z}\oplus T{\bf Z}$), located at $v = b/2$.
We may also find $\zeta_b$ with another representation:
\eqn\cotanz{\zeta_b(v) = \sum_{m \in {\bf Z}}
e^{-{\rm i}\pi b m}\,\pi\,{\rm cotan}\,\pi(v + mT)\ ,}
where $\sum_{m \in {\bf Z}}\cdots$ has to be understood as
$\lim_{M \rightarrow \infty} \left(\sum_{m = -M}^{M}\cdots\right)$.

Let us introduce coefficients $C_0$ and $C_1$ such that
\eqn\zetaexp{\zeta_{b}(v) = {1 \over v} + C_0 +
C_1\,v + O(v^2),\qquad v \rightarrow 0\ .}
In other words:
\eqn\coefs{C_0 = (\ln \vartheta_1)'(-b/2|T),\qquad C_1 =
{1 \over 2}\,{\vartheta_1''(-b/2|T) \over \vartheta_1(-b/2|T)} -
{1 \over 6}\,{\vartheta_1'''(0|T) \over \vartheta_1'(0|T)}\ .}
Several differential equations can be derived for $\zeta_b(v)$. They are all
based on the fact that $\zeta_b^{(j)}(v)/\zeta_b(v)$ is $1$ and $T$ periodic,
thus can be expressed just by matching the divergent behavior at the poles
with help of Weierstra{\ss}{} function and its derivatives.
\eqn\diffz{\eqalign{\zeta_b'(v) & =  {1 \over 2}\,{\wp'(v) + \wp'(b/2)
\over \wp(v) - \wp(b/2)}\,\zeta_b(v) \cr
\zeta_b''(v) & =  2C_0\zeta'_b(v) + 2(\wp(v) - C_1)\zeta_b(v) \ . \cr}}
By the same method, one finds a ``mirror relation'':
\eqn\mirr{\zeta_b(v)\zeta_b(-v) = \wp(b/2) - \wp(v)}
which shows that $\wp(b/2) = C_0^2 - 2C_1$. As a consequence, we mention that
${\tilde \zeta_b(v)} = e^{-C_0 v}\zeta_b(v)$ satisfies the spin $1$
Lam\'e differential equation with spectral parameter $b/2$:
\eqn\lame{{\tilde \zeta}_b''(v) - 2\wp(v){\tilde \zeta}_b(v) =
\wp(b/2){\tilde \zeta}_b(v)\ .}

\listrefs

\end